# "Open Access" to Monopoly Cable Platforms Versus "Direct Access" To Competitive International Telecommunications Satellite Facilities: A Study In Contrasts.

by

**Ken Katkin**
**Assistant Professor of Law**
**Salmon P. Chase College of Law**
**Northern Kentucky University**
**(859) 572-5861**
**katkink@nku.edu**

# ABSTRACT

abstract
The FCC is now engaged in resolving whether to require cable system operators who provide cable modem service to residential users to furnish cable transmission capacity to unaffiliated Internet Service Providers.  To resolve this controversy, the FCC has expressed a desire "to develop an analytical approach that is, to the extent possible, consistent across multiple platforms."  This comment may have been intended to specifically highlight the fact that DSL service is currently subject to a panoply of access and unbundling requirements that do not now apply to cable modem service.  However, it can also be read more broadly to suggest that in a world of increasing technological convergence and increasing intermodal competition, a more universally consistent analytical approach is needed to resolve the many analogous disputes over competitive access to proprietary bottleneck facilities that arise in a broad range of communications contexts.

The issues raised by the current dispute over "cable open access" are substantially analogous to those raised in the longstanding dispute over "direct access" to the INTELSAT satellite system.  That dispute was resolved in 1999, when the FCC authorized unaffiliated competitors to obtain direct access to INTELSAT, on the grounds that such a policy would :  (1) encourage the widest possible deployment of communications facilities; (2) encourage competition among providers of communications service; and (3) benefit consumers by facilitating lower prices and more diverse service offerings.  Each of these arguments have been raised by proponents of cable open access.  In fact, without exception, these criteria each support the implementation of cable open access today at least as strongly as they supported the implementation of direct access to INTELSAT in 1999.  Accordingly, implementation of cable open access would be analytically consistent with the implementation of direct access to INTELSAT.  Conversely, an FCC decision *not* to implement cable open access would be analytically *inconsistent* (indeed, irreconcilable) its decision to impose INTELSAT direct access decision.

## TABLE OF CONTENTS

I.      INTRODUCTION AND SUMMARY

Almost as soon as high-speed Internet access became available to residential users via cable modem in 1998, a controversy erupted over whether the cable system operators who owned the HFC cables used to provide such access should be required by law to furnish cable transmission capacity to unaffiliated Internet Service Providers (ISPs).[1] Proponents of such a regulatory requirement refer to it as "cable open access," and assert that is needed in order to create or maintain conditions of competition in the residential ISP market.[2] Opponents, in contrast have asserted that such "forced access" would be unfair to incumbent cable operators, and would discourage future investment in the continued deployment of "last mile" cable facilities that connect residential users to the Internet.[3]

---

[1]     *See* Jim Chen, *The Authority To Regulate Broadband Internet Access Over Cable*, 16 Berkeley Tech. L.J. 677, 677 (2001) ("The regulation of cable- based platforms for high-speed access to the Internet has become the most controversial subject in communications law."). *Cf.* James B. Speta, *The Vertical Dimension of Cable Open Access*, 71 U. Colo. L. Rev. 975, 980 (2000) (the "cable open access" issue "seems to have been placed into public debate, if not to have been first born, upon the announcement of the AT&T/TCI merger [in 1998]").

[2]     *See, e.g.*, Mark A. Lemley & Lawrence Lessig, *The End of End-To-End: Preserving the Architecture of the Internet in the Broadband Era*, 48 UCLA L. Rev. 925 (2001); Mark A. Lemley & Lawrence Lessig, *Open Access to Cable Modems*, 22 Whittier L. Rev. 3 (2000). *See also* Jerry A. Hausman, J. Gregory Sidak, & Hal J. Singer, *Residential Demand For Broadband Telecommunications and Consumer Access To Unaffiliated Internet Content Providers*, 18 Yale J. On Reg. 129, 170-71 (2001) ("To remedy the risks of conduit and content discrimination, regulators should subject any pending mergers to an open access provision."); Steven A. Augustino, *The Cable Open Access Debate: The Case for a Wholesale Market*, 8 Geo. Mason L. Rev. 653, 655 (2000) (arguing that not just ISPs, but also telecommunications service providers, should enjoy a right of open access to residential cable transmission capacity); Marcus Maher, Comment, *Cable Internet Unbundling: Local Leadership In the Deployment of High-Speed Access*, 52 Fed. Comm. L.J. 211, 221-23 (1999). *See generally Notice of Inquiry Concerning High-Speed Access to Internet*, 15 FCC Rcd. 19287, ¶ 27 (2000) (defining "cable open access" and summarizing arguments for and against requiring it).

[3]     *See, e.g.*, John E. Lopatka & William H. Page, *Internet Regulation and Consumer Welfare: Innovation, Speculation, and Cable Bundling*, 52 Hastings L.J. 891 (2001); Julian Epstein, *A Lite Touch On Broadband: Achieving the Optimal Regulatory Efficiency in the Internet Broadband Market*, 38 Harv. J. on Legis. 37 (2001); James B. Speta, *Handicapping the Race for the Last Mile?: A Critique of Open Access Rules for Broadband Platforms*, 17 Yale J. on Reg. 39 (2000). The phrase "last mile" is a term of art that refers to "the communications links and related hardware that connect the premises with the rest of a telecommunications network, most notably between the home

Initially, much of the "cable open access" controversy revolved around the threshold question of whether existing law *already* required (or prohibited) such access.[4] On March 15, 2002, however, the Federal Communications Commission substantially resolved this threshold question, by formally classifying the provision of high-speed Internet access to residential users via cable modems as an "information service," and *not* as a "cable service" or a "telecommunications service."[5] By choosing to classify cable modem services as "information services," the FCC maximized its own continuing

---

or small business and the set of interlinked data networks that make up the Internet." Committee on Broadband Last Mile Technology, Computer Science and Telecommunications Board, National Research Council, *Broadband: Bringing Home the Bits* 5 (2002); *see also id.* at 45 ("The link between the [Internet] point of presence and the customer—using either existing communications infrastructure or new facilities—is frequently referred to as the 'last mile' because it represents a bottleneck that constrains the benefits that the consumer gets from the rest of a network, which is literally at some distance."). In the context of residential cable modem service, the "last mile" facility is the hybrid fiber-coax (HFC) cable that delivers access to the Internet into the user's home. In the same context, a "facilities-based provider" is a provider that uses its own proprietary "last mile" facilities to deliver high-speed Internet access to users.

[4] *See generally* Jim Chen, *The Authority To Regulate Broadband Internet Access Over Cable*, 16 Berkeley Tech. L.J. 677, 704, 712 (2001) (concluding that "state and local authorities [are barred] from demanding open access," but that "the FCC may issue an open access rule for cable broadband platforms under any of several general grants of rulemaking power"); *See also* Lawrence A. Sullivan, *Is Competition Policy Possible In High Tech Markets? An Inquiry Into Antitrust, Intellectual Property, and Broadband Regulation As Applied To "The New Economy,"* 52 Case W. Res. L. Rev. 41, 82-86 (2001) (discussing FCC's authority to impose cable open access requirements); Barbara S. Esbin & Gary S. Lutzker, *Poles, Holes and Cable Open Access: Where The Global Information Superhighway Meets The Local Right-Of-Way*, 10 CommLaw Conspectus 23 (2001) (same); Christopher E. Duffy, Note, *The Statutory Classification of Cable-Delivered Internet Service*, 100 Colum. L. Rev. 1251, 1262 (2000). *Cf.* Raymond Shih Ray Ku, *Open Internet Access and Freedom of Speech: A First Amendment Catch-22*, 75 Tul. L. Rev. 87 (2000) (concluding that mandatory open access requirements would be unconstitutional).

[5] *See In re Inquiry Concerning High-Speed Access to the Internet Over Cable and Other Facilities, Declaratory Ruling & NPRM*, FCC 02-77, GEN Docket No. 00-185, 2002 WL 407567, ¶ 7 (rel. Mar. 15, 2002) ("*Cable Modem Order & NPRM*") (holding that "cable modem service, as it is currently offered, is properly classified as an interstate information service, not as a cable service, and [it involves] . . . no separate offering of telecommunications service."), *petitions for review pending sub nom.*, *EarthLink, Inc. v. FCC*, Docket No. 02-1097 (D.C. Cir. filed Mar. 26, 1997).

discretion to adopt—or not to adopt—"cable open access" requirements.[6] To facilitate its exercise of this discretion, the FCC simultaneously launched a new rulemaking proceeding to consider the merits of "whether it is necessary or appropriate at this time to require that cable operators provide unaffiliated ISPs with the right to access cable modem service customers directly. . . ."[7]

At present, wireline telephone local exchange carriers ("LECs") who provide residential high-speed Internet services via "Digital Subscriber Line" ("DSL") are subject to access and unbundling requirements that do not apply to cable operators who provide fungible services.[8] Accordingly, in arguing for (or against) "cable open access," commentators have frequently drawn analogies (or distinctions) between residential high-speed cable modem service and residential high-speed DSL service.[9] Cognizant of such

---

[6]  *See* Part V.B, *infra* (discussing regulatory consequences of classification of cable modem services as "information services"). Of course, the FCC's classification of cable modem services as "information services" cannot provide the agency with authority to implement cable open access if open access is unconstitutional, as some have alleged. *Cf.* Raymond Shih Ray Ku, *Open Internet Access and Freedom of Speech: A First Amendment Catch-22*, 75 Tul. L. Rev. 87 (2000) (asserting that mandatory open access requirements would be unconstitutional).

[7]  *In re Inquiry Concerning High-Speed Access to the Internet Over Cable and Other Facilities, Declaratory Ruling & NPRM*, FCC 02-77, GEN Docket No. 00-185, 2002 WL 407567, ¶ 72 (rel. Mar. 15, 2002), *petitions for review pending sub nom.*, *EarthLink, Inc. v. FCC*, Docket No. 02-1097 (D.C. Cir. filed Mar. 26, 1997). In this new rulemaking proceeding, the Commission continued to solicit commentary concerning the Commission's legal authority to implement "cable open access." *See id.* ¶¶ 72, 79 (soliciting further comment on the scope of the FCC's statutory authority to implement direct access); *id.* ¶¶ 80-82 (soliciting comment on possible constitutional limitations on the FCC's authority to implement direct access). The Commission also made clear, however, that it is now ready to consider the substantive merits of implementing such a policy *See id.* ¶ 73 (setting forth substantive considerations which the FCC seeks comment on).

[8]  *See* notes __, *infra*; *see also In re Notice of Inquiry Concerning High-Speed Access to Internet, Notice of Inquiry*, 15 FCC Rcd. 19287, ¶ 43 (2000) (as common carriers, most wireline LECs "must allow ISPs to purchase basic transmission services on a nondiscriminatory basis. As a result, end users are typically given a choice of ISPs, which could be accessed over the telephone network. Cable operators . . . do not currently operate pursuant to rules requiring end user ISP choice."). For a review of the historical roots of the present divergence between the regulatory paradigms now applied to DSL and cable modem service, *see* Rosemary Harold, **Cable-Based Internet Access: Exorcising the Ghosts of "Legacy" Regulation,** 28 N. Ky. L. Rev 721 (2001).

[9]  *See generally, e.g.*, Comments of SBC Corp. and BellSouth Corp *filed in* FCC GEN Docket No. 00-185 (filed Dec. 1, 2000) (arguing that because cable modems and

analogies, the FCC has now resolved to address whether it is appropriate for cable and DSL Internet access to continue to be subject to substantially dissimilar regulatory regimes.[10]

When it framed this question, the Commission stated that it will "strive to develop an analytical approach that is, to the extent possible, consistent across multiple platforms."[11] At a minimum, the Commission has thus identified the establishment of regulatory parity between cable modem and DSL service as a desirable policy objective, albeit one that might not necessarily trump other policy objectives or statutory constraints. More generally, however, the Commission's statement might reflect a broader desire to establish a more consistent analytical approach to resolving the disputes over competitive

---

DSL lines provide fungible high-speed Internet access services to residential users, the two technologies should be subject to the same regulatory requirements). *Cf.* James B. Speta, *Handicapping the Race for the Last Mile?: A Critique of Open Access Rules for Broadband Platforms*, 17 Yale J. on Reg. 39, 42 (2000) ("On one end of the spectrum, incumbent local telephone companies are currently subject not only to a significant remnant of traditional public utility regulation, but also the new interconnection, unbundling, and cooperation duties imposed by the 1996 [Telecommunications] Act. . . . At the other end, access providers that do not use any of the existing telephone plant (such as . . . cable television companies . . .) may not be required even to interconnect their facilities with those of other networks.").

[10] *See In re Inquiry Concerning High-Speed Access to the Internet Over Cable and Other Facilities, Declaratory Ruling & NPRM*, FCC 02-77, GEN Docket No. 00-185, 2002 WL 407567, ¶ 6 (rel. Mar. 15, 2002) ("[I]n this proceeding, as well as in a related proceeding concerning broadband access to the Internet over domestic wireline facilities, we seek to create a rational framework for the regulation of competing services that are provided via different technologies and network architectures. We recognize that residential high-speed access to the Internet is evolving over multiple electronic platforms, including wireline, cable, terrestrial wireless and satellite.") (footnote omitted), *petitions for review pending sub nom.*, *EarthLink, Inc. v. FCC*, Docket No. 02-1097 (D.C. Cir. filed Mar. 26, 1997). The "related proceeding concerning broadband access to the Internet over domestic wireline facilities" referred to in the *Cable Modem Order & NPRM* proposes to classify high-speed residential DSL service as an "information service" subject to substantially similar regulatory requirements as cable modem service. *See In re Appropriate Framework for Broadband Access to the Internet Over Wireline Facilities*, *Universal Service Obligations of Broadband Providers*, *Notice of Proposed Rulemaking*, FCC 02-42, CC Docket No. 02-33, 2002 WL 252714 ("*Wireline Broadband NPRM*") (rel. Feb. 15, 2002), *reprinted in* 67 Fed. Reg. 9232 (Feb. 28, 2002).

[11] *In re Inquiry Concerning High-Speed Access to the Internet Over Cable and Other Facilities, Declaratory Ruling & NPRM*, FCC 02-77, GEN Docket No. 00-185, 2002 WL 407567, ¶ 73 (rel. Mar. 15, 2002), *petitions for review pending sub nom.*, *EarthLink, Inc. v. FCC*, Docket No. 02-1097 (D.C. Cir. filed Mar. 26, 1997).

access to proprietary bottleneck facilities which recur in a broad range of communications contexts. If so, then the FCC might wish to consider harmonizing its analytical approach to the "cable open access" debate not only with its analytical approach to DSL access requirements, but also with the approach that it recently employed to resolve another longstanding dispute which also concerned competitors' demands for a right of unbundled access to separate components of proprietary communications facilities.

Specifically, in 1999, the FCC adopted a rule entitling unaffiliated international telecommunications carriers and users to obtain "direct access" at wholesale rates to the raw satellite transmission capacity of the INTELSAT satellite system, without being required to purchase any additional "bundled" communications services from INTELSAT's U.S. affiliate, COMSAT.[12] In authorizing "direct access" to INTELSAT, the FCC confronted and resolved several issues of law and policy that are analytically analogous to those raised by the current dispute over whether unaffiliated ISPs should be entitled to obtain "open access" at wholesale rates to the raw cable transmission capacity that is used to provide retail residential high-speed Internet service. As discussed herein, the Commission's decision to implement direct access to INTELSAT was predicated on its findings that adopting such a rule would : (1) encourage the widest possible deployment of communications facilities; (2) encourage competition among providers of communications service; (3) maintain regulatory neutrality among competing technologies that provide similar services; and (4) remove regulation where the public interest is served by that action.[13] These are substantially the same policy benefits that proponents claim would derive from a rule authorizing "cable open access" rule.[14] Accordingly, this Paper contends that the FCC's stated goal of harmonization of analytical approach would militate in favor of implementing cable open access or, alternatively, possibly repealing the rule requiring direct access to INTELSAT.[15]

---

[12] *See Direct Access to the INTELSAT System*, 14 FCC Rcd. 15703 (1999). INTELSAT is "a 143-member intergovernmental organization created by international agreement." *In re INTELSAT L.L.C.*, 15 FCC Rcd. 15460, ¶ 5 (2000), *recon. denied*, 15 FCC Rcd. 28234 (2000) (citing Agreement Relating to the International Telecommunications Satellite Organization "INTELSAT," done Feb. 12, 1973, 23 U.S.T. 3813 ("INTELSAT Agreement") and Operating Agreement Relating to the International Telecommunications Satellite Organization, "INTELSAT," done Aug. 20, 1971, 23 U.S.T. 4091 ("INTELSAT Operating Agreement")). Until 2001, the INTELSAT treaty organization owned and operated a global fleet of geostationary commercial communications satellites over which much of the world's international telephone, video, data, Internet, and other communications were transmitted. *Id.* ¶¶ 5-7. On July 18, 2001, INTELSAT's business operations were privatized "into a corporate holding company structure." *Intelsat L.L.C.*, 16 FCC Rcd. 18185, ¶ 1 (2001). Today, INTELSAT's former satellite fleet is operated by Intelsat Ltd., a Delaware corporation that "is a subsidiary within that privatized structure and the U.S. licensee for operation of existing and planned satellites in the C-band and Ku- band." *Id.* For more on INTELSAT, see Subparts II.B, III.B, and V.A, *infra*.

[13] *See* Subpart VI, *supra*.

[14] *See id.*

Part II of this Paper describes the parallel development in the 1960s of cable television and international satellite telecommunications, both of which developed under conditions of regulated "natural monopoly." Part III describes the onset in the 1980s of competition against the incumbents in both industries, and the regulatory readjustments that were made to foster or to accommodate such competition. Part IV describes the rise of residential Internet access and the separate regulatory frameworks that characterize the low-speed and high-speed residential ISP markets. Part V recounts the issues raised in the separate debates over "cable open access" and "INTELSAT direct access." On an issue-by-issue basis, Part VI compares the merits of the case for "cable open access" with those of the case for "INTELSAT direct access." Without exception, this Paper concludes that the stated criteria underlying the FCC's decision to implement direct access to INTELSAT in 1999 each would provide at least as strong a basis for implementing cable open access today. Accordingly, implementation of cable open access would be analytically consistent with the Commission's stated reasons for implementing direct access to INTELSAT. Conversely, an FCC decision *not* to implement cable open access would be analytically *inconsistent* (indeed, irreconcilable) with the INTELSAT direct access decision.

## II.   Cable TV and INTELSAT Satellites:  The Monopoly Years.

In the 1960s, America witnessed the rise of two new means of communicating information: cable television and geostationary international telecommunications satellites ("GEOs" or "satellites").[16] In certain respects, the business and legal arrangements that

---

[15]   On first glance, the FCC might appear to now lack authority to repeal its rule requiring direct access to INTELSAT. In 1999, the FCC implemented direct access to INTELSAT in a rulemaking proceeding. *Direct Access to the INTELSAT System*, 14 FCC Rcd. 15703 (1999). Six months later, however, Congress codified the FCC's direct access rule, seemingly immunizing the rule against FCC repeal. *See* 47 U.S.C. § 765(a) (enacted Mar. 17, 2000). In July 2001, however, as discussed at note __, *supra*, INTELSAT's satellites were transferred to Intelsat Ltd., an Delaware corporation defined in the ORBIT Act as a "successor entity." *Compare* 47 U.S.C. § 769(a)(1) (defining "INTELSAT") *with* 47 U.S.C. § 769(a)(7) (defining "successor entity"). Because the statutory "direct access" requirement set forth at 47 U.S.C. § 765(a) applies only to the intergovernmental treaty organization "INTELSAT" and not to any private "successor entit[ies]," the statute no longer applies and the FCC's authority to repeal its direct access rule has been restored.

[16]   For a collection of essays on the history of cable television in the United States, *see Milestones: A History of Cable Television* (Priscilla Walker & Matt Stump eds. 1998). For a history of international telecommunications satellites, see Charles H Kennedy & M. Veronica Pastor, *An Introduction to International Telecommunications*
<space count="40"/>(continued . . . )

<space count="40"/>6

developed around these new communications technologies were (and remain) quite different from one another. Cable television system operators, for example, directly serve the residential retail consumers that are cable's end users. International GEOs, in contrast, do not serve any retail consumers directly, but instead provide transmission capacity to the U.S. telecommunications carriers and other users (such as television networks and ISPs) who use that capacity to provide international calling, data, or video services to the public. Moreover, although subject to federal regulation in some limited respects, cable television is (and has always been) regulated primarily by the State and local authorities that must issue and renew cable franchises.[17] GEOs, in contrast, are located 22,300 miles above the earth—beyond the regulatory reach of any State or local authority. Accordingly, GEOs are regulated in some respects by the federal government,[18] and in other respects by international treaty organizations such as the International Telecommunications Union (ITU), an agency of the United Nations.[19]

In significant respects, however, the economics of cable television resembles that of international satellite communications. Both industries—like local telephony—operate under conditions where the fixed cost of entry (constructing facilities) is unusually high in comparison with the low marginal cost of providing additional service once a facility has been built.[20] Primarily because of the economic disincentive to subsequent entry that

---

*Law* 50-97 (1996).

[17]  *See, e.g.*, *Community Communications Co., Inc. v. City of Boulder*, 660 F.2d 1370, 1377-78 (10th Cir. 1981), *cert. dismissed*, 456 U.S. 1001 (1982) ("[A] cable operator must lay the means of his medium underground or string it across poles in order to deliver his message. Obviously, this manner of using the public domain entails significant disruption, especially to streets, alleys, and other public ways. Some form of permission from the [local] government must, by necessity, precede such disruptive use of the public domain. We do not see how it could be otherwise. A city needs control over the number of times its citizens must bear the inconvenience of having its streets dug up and the best times for it to occur.") (footnotes and citations omitted).

[18]  *See, e.g.*, 47 U.S.C. §§ 701-69 (setting forth federal statutes applicable to international telecommunications satellites that serve the United States); 47 C.F.R. Part 25 (setting forth FCC regulations applicable to such satellites).

[19]  *See* Charles H Kennedy & M. Veronica Pastor, *An Introduction to International Telecommunications Law* 52-58 (1996).

[20]  *See, e.g. Omega Satellite Prods. Co. v. City of Indianapolis*, 694 F.2d 119, 126 (7th Cir. 1982) (Posner, J.) ("The cost of the cable grid appears to be the biggest cost of a cable television system and to be largely invariant to the number of subscribers the system has. . . . [O]nce the grid is in place—once every major street has a cable running above or below it that can be hooked up to the individual residences along the street—the cost of adding another subscriber probably is small. If so, the average cost of cable television would be minimized by having a single company in any given geographical

(continued . . .)

normally obtains under such conditions, both the international satellite telecommunications industry and the cable television industry were long believed to be inherently subject to "natural monopolies."[21] This economic theory of "natural monopoly" was reflected in the respective regulatory regimes that evolved around the two industries: in both cases, monopolies were originally protected, while rates were regulated.

### A. The Development of Cable Television as a "Natural Monopoly."

Historically, local government regulation of cable television systems "has been premised upon cable companies' need to use public streets and rights of way to lay or string their cable."[22] Specifically, local governments have alleged that there is "a sheer limit on the number of cables that can be strung on existing telephone poles."[23] This "sheer limit" has been analogized to the physical "spectrum scarcity" that has served as a

---

area; for if there is more than one company and therefore more than one grid, the cost of each grid will be spread over a smaller number of subscribers, and the average cost per subscriber, and hence price, will be higher"); *see also* pages __, *infra* (discussing the reasons why Congress in the early 1960s believed that an international satellite telecommunications system would constitute a natural monopoly).

[21] *See, e.g. Omega Satellite Prods. Co. v. City of Indianapolis*, 694 F.2d 119, 126 (7th Cir. 1982) (Posner, J.) (presenting economic argument for why cable television service has been thought to be a "natural monopoly"); *Regulatory Policies And International Telecommunications*, 2 FCC Rcd. 1022, ¶ 18 (1987) (surveying reasons why international telecommunications service was historically thought to be a "natural monopoly"), *modified in other respects*, 4 FCC Rcd. 7387 (1988), *and modified on further recon. in other respects,* 7 FCC Rcd. 1715 (1992). In economic theory, a "natural monopoly" is defined as a firm that "can exist with decreasing returns if any specified required rate of output can be supplied most economically by a single firm or single system." Thomas Hazlett, *The Curious Evolution of Natural Monopoly Theory*, in *Unnatural Monopolies* 1, 15 (Robert W. Poole, Jr. ed. 1985); *accord* William W. Sharkey, *The Theory of Natural Monopoly* 4 (1982); *see also McGraw-Hill Dictionary of Modern Economics* 394 (2d ed. 1973) (defining "natural monopoly" as "a natural condition that makes the optimum size of the firm so large in relation to the market that there is room for only one firm.").

[22] *Community Communications Co., Inc. v. City of Boulder*, 660 F.2d 1370, 1374 (10th Cir. 1981), *cert. dismissed*, 456 U.S. 1001 (1982). "Local regulation has commonly taken the form of licensing or franchising cable companies." *Id.* (citing Albert, *The Federal and Local Regulation of Cable Television*, 48 U. Colo. L. Rev. 501, 508-13 (1977)).

[23] *Id.* at 1378.

justification for otherwise anomalous federal regulation of the use of the broadcast spectrum.[24]

In addition, however, local governments have also advanced as a second basis for government regulation of cable entry. Under the hybrid doctrine of "medium scarcity," local government regulators have asserted "that there are physical *and economic* limitations on the number of cable systems that can practicably operate in a given geographic area."[25] That is, even though the "sheer [physical] limit on the number of cables that can be strung on existing telephone poles" might in some cases accommodate more than one such cable, some local governments have nonetheless also asserted "that cable broadcasting is a monopolistic industry because it is not economically viable for more than one cable company to operate in any given geographic area."[26] Together, proponents of "medium scarcity" theory have contended, cable's dual physical *and* economic limitations give the industry "the character of a natural monopoly and thus make the cable broadcasting medium 'scarce' in much the same way that the finiteness of the electromagnetic spectrum makes wireless broadcasting a medium of essentially limited access."[27]

Based on the theory of "medium scarcity," many local governments believed that it would be contrary to the public interest to permit competing cable systems to "overbuild"[28] one another within the same community. Judge Richard Posner summarized (and endorsed) this view as follows:

> You can start with a competitive free-for-all—different cable
> television systems frantically building out their grids and signing up subscribers in an effort to bring down their average costs faster than their rivals—but eventually there will be only a single company, because until a company serves the whole market it will have an incentive to keep expanding in order to lower its average costs. In the interim there may be wasteful duplication of facilities. This duplication may lead not only to higher prices to cable television subscribers, at least in the short run, but also to higher costs to other users of the public ways, who must compete with the cable television companies for access to them. An alternative procedure is to pick the most efficient competitor at the outset, give him a monopoly, and extract from him in exchange a commitment to provide reasonable service at reasonable rates.[29]

---

[24] *Id.*; *cf. Red Lion v. FCC*, 395 U.S. 397 (1969) (holding that but for "spectrum scarcity," the existing system of broadcast licensing under the Communications Act of 1934 might otherwise violate the First Amendment).

[25] *Community Communications Co., Inc.*, 660 F.2d at 1378 (emphasis added).

[26] *Id.*

[27] *Id.*; see also *Omega Satellite Prods. Co. v. City of Indianapolis*, 694 F.2d 119, 126 (7th Cir. 1982) (Posner, J.) (cable television's unique characteristics may constitute "what economists call a 'natural monopoly,' wherein the benefits, and indeed the very possibility, of competition are limited.").

[28] Cable "overbuilding" is defined as occurring "when two or more wireline cable systems directly compete for subscribers in a local delivery market." *Notice of Inquiry Concerning Deployment of Advanced Telecommunications Capability to All Americans in a Reasonable and Timely Fashion*, 15 FCC Rcd. 16641, ¶ 45 n.66 (2000) (citation omitted). Overbuilt cable systems would thus pose facilities-based intramodal competition to one another.

For these or similar reasons, many local governments from the 1960s through the 1980s sought to award, and protect, a monopoly franchise in cable service in their communities.

### B. The Development of the INTELSAT Satellite System as a "Natural Monopoly."

In 1961, President John F. Kennedy called upon the United States to spearhead a cooperative international effort to develop and operate the world's first global satellite communications system.[30] President Kennedy proposed that the satellite system should provide global coverage, with nondiscriminatory access, at the earliest practicable date.[31]

Beginning in 1961 and through the summer of 1962, Congress debated how best to achieve President Kennedy's ambitious goal.[32] Some legislators submitted proposals urging that the government build, finance, and operate the proposed satellite system.[33] Others favored allowing a consortium of existing U.S. telephone and telegraph companies to do the job.[34] In contrast, the Kennedy Administration proposed to charter

---

[29] *Omega Satellite Prods. Co.*, 694 F.2d at 126.

[30] *See* Statement of the President on Communications Policy (July 24, 1961), *appended to* S. Rep. No. 87-1584, at 25 (1962), *reprinted in* 1962 U.S.C.C.A.N. 2269, 2287.

[31] *Id.*; *see also* U.N. General Assembly Resolution 1721 (XVI), Part D (1962) (same).

[32] *See* Charles H Kennedy & M. Veronica Pastor, *An Introduction to International Telecommunications Law* 65 (1996) (summarizing Congressional debate).

[33] *See, e.g.,* S. 2890, 87th Cong. (1962).

[34] *See, e.g.*, S. 2650, 87th Cong. (1962) (proposing carrier joint venture).

a new, private corporation to build and operate the U.S. portion of the proposed international system.[35]

One key objective underlying the Administration's proposal to charter a new private corporation was to facilitate investment of private shareholder capital to finance, build, manage, and operate the U.S. portion of the proposed global satellite system. The creation of a new private corporation ensured that neither the taxpayers generally,[36] nor the ratepaying customers of the incumbent telephone and telegraph carriers, would be compelled to bear the risk that the satellite system would not succeed. Instead, only those persons who voluntarily chose to invest in the new corporation would incur the risk associated with the corporation's obligation to "plan, initiate, construct . . . and operate" the satellite system.[37] Moreover, ownership of the new corporation would be structured "in a manner to encourage the widest distribution to the American public."[38]

Congress in 1962 believed that private investment in the new corporation would entail unusually high risk.[39] Accordingly, Congress recognized that "in a large measure the investors who would invest in the corporation will rely on the action of the Congress today to see to it that [the corporation] has a means to . . . make adequate earnings to justify their investment and their confidence in [the corporation]."[40] Thus, to enable investors to earn an adequate return if the satellite system proved successful, President Kennedy and his Congressional supporters favored granting the new corporation an exclusive right to furnish the channels of communication of the system to carriers and users located in the United States.[41]

---

[35] *See, e.g.*, *Communications Satellite Legislation: Hearings on S. 2650 and S. 2814 Before the Senate Comm. on Aeronautical and Space Science,* 87th Cong. 388 (1962) (testimony of Deputy Attorney Gen. Nicholas deB. Katzenbach) (articulating the Administration's preference "that there would be one corporation engaged in the transmission of messages by satellite, performing services for all authorized communications carriers in this country. . . .").

[36] *See* 108 Cong. Rec. S16694 (daily ed. Aug. 16, 1962) (statement of Sen. Humphrey) (supporting the bill primarily because "the cost of this communications satellite system will be privately financed rather than have it come out of the Federal budget. . . . What we are attempting to do is to adapt the resources of the country and to bring in the capital through other means than taxation.").

[37] 47 U.S.C. § 735(a)(1).

[38] 47 U.S.C. § 734(a).

[39] *See, e.g.*, S. Rep. No. 87-1584, at 11 (1962), *reprinted in* 1962 U.S.C.C.A.N. at 2273 ("it should be recognized that purchase of [the new corporation's] stock will be speculative"); 108 Cong. Rec. H7502 (daily ed. May 2, 1962) (statement of Rep. Harris) ("it is most important that the membership of the House and the investing public be completely aware of [the] unusual circumstances which render investments in this new corporation highly speculative").

[40] 108 Cong. Rec. H7525 (daily ed. May 2, 1962) (statement of Rep. Dingell).

Another key objective underlying the Kennedy Administration's proposal was to interpose the new corporation as a counterweight to AT&T, which was then a monopoly carrier that had already made substantial investments in transoceanic submarine cable, as well as nascent investments in satellite technology. The Administration sought to prevent AT&T from using its substantial market power to dominate any consortium of existing carriers, and thereby possibly thwarting the new satellite system's development or driving weaker rivals from the field.[42] The Administration's desire to counterbalance AT&T's dominance was given additional effect by structural safeguards designed "[t]o prevent any single interest or group of interests from dominating the activities of the [proposed new] corporation. . . ."[43]

The Administration's proposal to create a new private monopoly was not without its critics, in Congress and elsewhere. Congressional critics of the proposal complained that U.S. communications carriers carrying international traffic—and foreign carriers carrying U.S.-bound traffic—should not be required to "deal[] with a third party—the satellite corporation," in order to access the facilities of the satellite system.[44] Nonetheless, after extensive debate, Congress enacted the essence of the Kennedy Administration proposal as the Communications Satellite Act of 1962 ("Satellite Act").[45]

---

[41]　*See, e.g.*, S. Rep. No. 87-1584, at 28 (1962), *reprinted* in 1962 U.S.C.C.A.N. 2269, 2289 (statement of President Kennedy) (characterizing COMSAT as "by nature a Government-created monopoly. . . ."); *Communications Satellite Act of 1962*, S. Rep. No. 87-1319, at 2 (1962) ("The President indicated in his statement accompanying the proposal that such a corporation was by nature a Government-created monopoly. . . ."); *see also* 108 Cong. Rec. H7505 (daily ed. May 2, 1962) (statement of Rep. Cellar) (noting that in the Satellite Act, "we are creating here a private *monopoly*.") (emphasis added). *See generally COMSAT Study*, 77 F.C.C. 2d 564, 587 ¶ 54 (1980) (the Satellite Act "creates a single entity in the form of a private corporation to carry out its objectives and purposes … [and] endows [COMSAT] with extraordinary powers and privileges to carry out its mission, including monopoly status in the provision of services via the satellite system to authorized U.S. users."); *accord id.* at 591 (same).

[42]　*See, e.g., Communications Satellites: Hearings on HR. 10115 and H.R. 10138 Before the House Committee on Interstate and Foreign Commerce*, 87th Cong. 2d Sess., pt. 2, at 563 (1962) (statement of Att'y Gen. Robert F. Kennedy) (contending that an AT&T-led carrier consortium would "not have the same interest in promoting and guaranteeing nondiscriminatory use and equitable access to the system by competitors as would an independent corporation"); *see also* Delbert D. Smith, *Communication Via Satellite* 74 (1976) ("Concern over the possibility of an AT&T monopoly was one factor that prompted [Congress to] reorient[] . . . the direction that commercialization seemed to be following").

[43]　S. Rep. No. 87-1584, at 11 (1962), *reprinted in* 1962 U.S.C.C.A.N. 2269, 2272.

The opening provisions of the Satellite Act provided that "United States participation in the global system shall be in the form of *a* private corporation."[46] That corporation was to operate "for profit."[47] The Act authorized the corporation (COMSAT) to "own" and "operate" the U.S. portion of the proposed global satellite system and to "furnish, for hire, channels of communication" to U.S. carriers and other U.S. users of that system.[48] The Act also directed the FCC to insure that all of COMSAT's authorized carrier-customers have "nondiscriminatory use of" and "equitable access to" the system under just and reasonable rates, terms, and conditions.[49]

Consistent with the Kennedy proposal, a "basic tenet of the Satellite Act . . . [was that at] least insofar as international common carrier communications services are concerned, Comsat is given a virtual *statutory monopoly position* with respect to the operation of the space segment of the commercial communications satellite system."[50]

---

[44] 108 Cong. Rec. H7701 (daily ed. May 3, 1962) (statement of Rep. Hemphill). *See also* S. Rep. No. 87-1584, at 51 (1962), *reprinted* in 1962 U.S.C.C.A.N. 2269, 2309 (minority views) (opposing proposal to "create a private corporation that would own and operate the U.S. portion of a worldwide satellite communications system" on ground that "[t]his corporation would be a Government-created monopoly."); *see also, e.g.*, 108 Cong. Rec. H7515 (daily ed. May 2, 1962) (statement of Rep. Kowalski) ("Let us make no mistake about the bill before us—it proposes to place in private hands a Government-created monopoly. . . .").

[45] Communications Satellite Act of 1962, Pub. L. No. 87-624, 76 Stat. 425 (1962), *codified as amended at* 47 U.S.C. §§ 701-769 ("Satellite Act").

[46] 47 U.S.C. § 701(c) (emphasis added). The corporation that was formed pursuant to the Act was named "Communications Satellite Corporation," or "COMSAT." Later, the corporation formally changed its named to the Comsat Corporation.

[47] 47 U.S.C. § 731.

[48] 47 U.S.C. § 735(a)(1)-(2).

[49] 47 U.S.C. § 721(c)(2).

[50] *Authorized Entities & Authorized Users Under the Communications Satellite Act of 1962*, 4 F.C.C. 2d 421, 428 (1966) ("*Authorized User I*") (emphasis in original), *overruled by, Modification of Authorized User Policy*, 90 F.C.C. 2d 1394 (1982) ("*Authorized User II*"), *aff'd and remanded*, *ITT World Communications, Inc. v. FCC*, 725 F.2d 732 (D.C. Cir.1984).

Indeed, the FCC characterized "the general concept pervading the Satellite Act" as being the establishment of "Comsat as a monopoly (insofar as the space segment of international communications is concerned) and as primarily a carrier's carrier, created to provide at least the space segment of international communications as part of an improved global communications network consisting of all means of providing such communications services, so that lower rates should be possible to all the using public."[51]

After incorporating as a District of Columbia corporation in 1963, COMSAT immediately sought foreign partners willing to participate in establishing the proposed satellite system. In 1965, an *ad hoc* partnership led by COMSAT and involving 44 nations successfully launched into geostationary orbit the world's first commercial communications satellite, "Early Bird."[52] In 1971, after several more satellites had been launched by COMSAT-led *ad hoc* international partnerships,[53] 85 nations formed the permanent intergovernmental satellite organization, INTELSAT, "to continue and carry forward on a definitive basis the design, development, construction, establishment, operation and maintenance of the space segment of the global commercial telecommunications satellite system. . . ."[54] By virtue of its statutory role under the 1962 Satellite Act, COMSAT was designated as "the U.S. Signatory to the Operating Agreement of INTELSAT."[55]

---

[51] *Id.* at 430; *see also* Charles H Kennedy & M. Veronica Pastor, *An Introduction to International Telecommunications Law* 74 (1996) (The Satellite Act "created COMSAT as a corporation whose sole purpose [was] to act as the intermediary between INTELSAT and the U.S. international carriers, *and gave it a perennial monopoly*.") (emphasis added).

[52] *See Communications Satellite Corporation*, 5 Rad. Reg. 2d (P&F) 369, 371 (1965).

[53] *See COMSAT Study*, 77 F.C.C. 2d 564, ¶¶ 63-65 (1980) (describing interim arrangements).

[54] *Agreement Relating to The International Telecommunications Satellite Organization "INTELSAT"* Art. II(a), done Aug. 20, 1971, 23 U.S.T. 3813, T.I.A.S. No. 7532 ("INTELSAT Agreement"); *see also id.*, 23 U.S.T. at 4066-4083 (listing the 85 nations that founded INTELSAT). By 2000, just before INTELSAT was privatized, the number of member nations that had become Signatories to the INTELSAT Agreement had risen to 144. *See* United States Department of State, *Treaties In Force: A List of Treaties and Other International Agreements of the United States in Force as of January 1, 2000* 457-58 (2000) (listing INTELSAT member nations), <http://www.state.gov/www/global/legal_affairs/tif_01e.pdf>.

[55] S. Rep. No. 95-1036, at 4 (1978), *reprinted in* 1978 U.S.C.C.A.N. 5272, 5275. *See also Senate Report on International Maritime Satellite Telecommunications Act*, S. Rep. No. 95-1036, at 15 (1978), 1978 U.S.C.C.A.N. at 5286 (J.A. 406) ("COMSAT was created by the Communications Satellite Act of 1962 to foster the establishment of a

(continued . . .)

In 1978, COMSAT was also designated as the U.S. Signatory to the International Maritime Satellite Organization "Inmarsat," a second international treaty organization modeled on INTELSAT, which was chartered in order to develop and operate a global maritime satellite telecommunications system whose facilities would serve the maritime commercial and safety needs of the United States and foreign countries.[56] As with INTELSAT, COMSAT was to enjoy an exclusive franchise in furnishing the satellite space segment capacity of the Inmarsat system for communications between ships on the ocean and the United States.[57]

### C. The Regulatory Responses To "Natural Monopoly" in the International Satellite Telecommunications and Cable Television Industries.

Throughout the 1960s and 1970s, both the international satellite telecommunications industry and the local cable television industry were considered by regulators to be "natural monopolies."[58] For this reason, substantially similar regulatory principles were brought to bear on these two seemingly unrelated industries.

---

global satellite system and represent the United States in the operation aspects of that system. COMSAT is *therefore* the U.S. Signatory to the Operating Agreement of INTELSAT.") (emphasis added). As U.S. "Signatory" to INTELSAT, COMSAT signed the INTELSAT Operating Agreement on behalf of the United States, and represents the United States within INTELSAT. *See* INTELSAT Agreement Art. I(g), 23 U.S.T. at 3816 (defining "Signatory"); *see also* 47 U.S.C. §§ 721(a), 735(a), 742; *Operating Agreement Relating to the International Telecommunications Satellite Organization "INTELSAT"*, Art. 2, done Aug. 20, 1971, 23 U.S.T. 4091 ("INTELSAT Operating Agreement") (setting forth rights and obligations of Signatories).

[56] *See* International Maritime Satellite Telecommunications Act § 503(a)(1), Pub. L. No. 95-564, 92 Stat. 2392 (1978).*codified as amended at* 47 U.S.C. § 752(a)(1) (designating COMSAT "as the sole operating entity of the United States for participation in INMARSAT, for the purpose of providing international maritime satellite telecommunications services.").

[57] *See COMSAT Study*, 77 F.C.C. 2d 564, ¶ 86 (1980) ("INMARSAT will not involve a competitive marketplace environment. Comsat is the sole U.S. provider of space segment capacity obtained from INMARSAT; it will interconnect on a participating carrier basis with authorized U.S. domestic or international carriers for the extension of maritime satellite services with the United States and beyond. . . . Comsat will receive and assemble all traffic for appropriate routing, either inbound or outbound.") (footnote and citation omitted).

[58] *See* Subpart II.A., *supra* (cable/MVPD); Subpart II.B, *supra* (international satellite telecommunications).

For example, as "natural monopolies," both the incumbent local cable TV system operators and COMSAT were shielded by law against intramodal competition, which was thought to be "wasteful." Thus, until the mid-1980s, incumbent cable operators generally were shielded by local franchising authorities against the development of competing "overbuilt" cable systems in the same local market.[59] Analogously, during the same period, COMSAT was shielded by the FCC against the development of separate geostationary satellite systems capable of competing against INTELSAT to provide U.S.-international communications services.[60]

Conversely, neither incumbent local cable operators nor COMSAT ever were shielded by law against intermodal competition. Cable TV, of course, always faced at least some competition from free over-the-air broadcast TV.[61] Moreover, no law shielded

---

[59] *See, e.g.*, *Central Telecommunications, Inc. v. TCI Cablevision, Inc.*, 800 F.2d 711, 713-18 (8th Cir. 1986); *Affiliated Capital Corp. v. City of Houston*, 735 F.2d 1555, 1563 (5th Cir.), *cert. denied*, 473 U.S. 1053 (1984); *Omega Satellite Prods. Co. v. City of Indianapolis*, 694 F.2d 119, 126 (7th Cir. 1982); *Community Communications Co. v. City of Boulder*, 660 F.2d 1370, 1378-80 (10th Cir. 1981), *cert. dismissed*, 456 U.S. 1001 (1982); *Lamb Enters. v. Toledo Blade Co.*, 461 F.2d 506, 511 (6th Cir. 1972); *accord* Kent D. Wakeford, Note, *Municipal Cable Franchising: An Unwarranted Intrusion Into Competitive Markets*, 69 S. Cal. L. Rev. 233, 246 (1995) ("Under the assumption that cities could only economically support one cable franchise, municipal authorities issued exclusive franchises in the form of local monopolies."). For critical discussions of the history of municipal regulation of cable entry, *see id.* at 246-71 and Thomas W. Hazlett, *Private Monopoly and the Public Interest: An Economic Analysis of the Cable Television Franchise*, 134 U. Pa. L. Rev. 1335, 1358-59 (1986) (discussing political and economic incentives for local public authorities to use the cable franchising process a vehicle "to create durable monopolistic profit opportunities" for their political supporters).

[60] The Satellite Act itself did not preclude the development of separate international satellite systems capable of competing against INTELSAT. *See* 47 U.S.C. § 701(d) ("It is not the intent of Congress by this chapter to preclude . . . the creation of additional communications satellite systems, if required to meet unique governmental needs or if otherwise required in the national interest."). Under the INTELSAT Agreement, however, the INTELSAT treaty organization was vested with a right to veto the development of any separate international satellite systems that would cause "economic harm" to INTELSAT. *Agreement Relating to the International Telecommunications Satellite Organization "INTELSAT,"* Art. XIV(d), done Aug. 20, 1971, 23 U.S.T. 3813 ("INTELSAT Agreement"). Although INTELSAT never actually exercised this veto, the initial opposition of the United States to the development of separate systems may have delayed the development of competing geostationary communications satellite systems from many years. *See* Charles H. Kennedy & M. Veronica Pastor, *An Introduction to International Telecommunications Law* 79-80 (1996).

[61] The cable industry has long argued "that it had no monopoly character because
(continued . . .)

cable against the competition that eventually arose from other "multichannel video programming distributors" ("MVPDs"), such as direct broadcast satellite ("DBS") and terrestrial "wireless cable" service.[62] Similarly, the INTELSAT system from its inception faced intermodal competition, albeit limited, in the international telecommunications market from terrestrial facilities such as transoceanic submarine cables and short wave radio.[63] For various reasons, however (probably including both technological considerations and the advantages of incumbency), intermodal competition made no substantial competitive inroads against either cable TV or INTELSAT until relatively recently.[64] Thus, both international satellite telecommunications and cable TV emerged

---

over-the-air television . . . and videodisks . . . offer[ed] competing sources of video entertainment. . . ." Ithiel de Sola Pool, *Technologies of Freedom* 173 (1983). In 1983, however, a leading commentator analogized such claims to those of "a railroad operator in the nineteenth century denying being a monopolist because anyone refused access to a train could use a horse and buggy." *Id.*

[62]   *See Satellite Television Corporation Authority to Construct an Experimental Direct Broadcast Satellite System*, 91 F.C.C. 2d 953 (1982) (approving the first applications to provide DBS service), *aff'd sub nom.*, *National Ass'n of Broadcasters v. FCC*, 740 F.2d 1190 (D.C. Cir. 1984); *see also Private Operational-Fixed Microwave Service, Multipoint Distribution Service, Multichannel Multipoint Distribution Service, Instructional Television Fixed Service, and Cable Television Relay Service*, 5 FCC Rcd. 6410 (1990) (authorizing new "wireless cable" service analogous to DBS service, but transmitting from terrestrial broadcast towers rather than from satellites), *corrected*, 5 FCC Rcd. 6666 (1990), *modified*, 6 FCC Rcd. 6764 (1991), *further modified*, 10 FCC Rcd. 7074 (1995), *and clarified on denial of recon.*, 11 FCC Rcd. 17003 (1996).

[63]   "Undersea cables provided the first mean of transmission for voice communication between the U.S. and foreign points. The first successful trans-Atlantic [telegraph] cable was run between Ireland and Newfoundland in 1865 and the first undersea telephone cable between Europe and the U.S. was TAT-1, laid in 1956." Alieen A. Pisciotta, Randall W. Sifers, Heather Wilson Aspasia, & A. Paroutsas, *Regulatory Considerations Affecting Investments In Global Satellite and Undersea Cable Systems*, 1263 PLI/Corp 399, 415 (2001). As early as 1870, transoceanic submarine cables had been used to connect five port cities in China with Shanghai. *See* Zhou He, *A History of Telecommunications In China: Development and Policy Implications*, in *Telecommunications and Development in China* 55, 57-60 (Paul S.N. Lee ed., 1997).

[64]   *See* Part III, *infra* (discussing the onset of intermodal competition). Even today, despite facing substantial facilities-based intermodal competition, cable continues to control a dominant share of the residential market for the delivery of multichannel video programming ("MVPD market"). *See In re Annual Assessment of Status of Competition in Market for Delivery of Video Programming*, *Eighth Annual Report*, 17 FCC Rcd. 1244, ¶¶ 7-8, 13 (2002) (noting that as of mid-2001, almost 69 million Americans were subscribing to cable television service, while only 19.3 million American homes were subscribing to non-cable MVPDs, most prominently Direct Broadcast Satellite (DBS)

as *de facto* monopolies, freed by law from facing any intramodal competition, and by circumstances from facing any intermodal competition.

Despite their respective monopolies, however, until recently neither INTELSAT nor incumbent cable systems were rendered subject to the panoply of unbundling or interconnection requirements that have long applied to telecommunications common carriers in possession of similar monopolies.[65] This result was not inevitable. In 1970, for example, several members of a commission on cable television established by the Sloan Foundation concluded that cable television systems should immediately be regulated as common carriers.[66] Moreover, a 1973 report by a Presidential Cabinet Committee on Cable Communications proposed an evolutionary approach under which cable operators would initially be permitted to select the programming carried over their cable systems, but would later be subject to common carrier regulation when cable penetration reached 50 percent of American homes.[67] Similarly, as discussed above, Congress in 1962 considered alternative approaches to organizing the proposed

---

service).

[65] Since the beginning of the century, all telecommunications common carriers have borne a duty to provide service to other carriers at just and nondiscriminatory rates. *See, e.g.*, *People* ex rel. *Western Union Tel. Co. v. Public Serv. Comm'n of N.Y.*, 129 N.E. 220, 222 (N.Y. 1920). This duty of physical interconnection at just and reasonable rates is reflected in the Communications Act of 1934, see 47 U.S.C. §§ 201, 203, 205, and was expanded considerably in the Telecommunications Act of 1996, 47 U.S.C. §§ 251-59. *See generally* Charles H. Kennedy, *An Introduction to U.S. Telecommunications Law* 35-36 (2d ed. 2001) (incumbent LECs "own and control an overwhelming preponderance of the switches, access lines, and other facilities through which non-ILEC service providers must reach their customers, and the ILECs' decisions concerning when and on what terms they will make their networks available to others can determine the extent—or indeed the very existence—of competition in a wide range of markets. For this reason Congress, the FCC, the state PUCs, and the antitrust courts have intervened extensively to dictate when, to whom, and on what terms the ILECs will offer wholesale services and facilities"); Howard Shelanski, *Regulating at the Technological Edge: New Challenges For the FCC*, 2000 L. Rev. Mich. St. U. Det. C.L. 3, 4 (2000) (explaining the scope of the LEC unbundling requirements set forth in the 1996 Act).

[66] Ithiel de Sola Pool, *Technologies of Freedom* 169 & n.33 (1983) (citing *On The Cable* (1971)).

[67] Ithiel de Sola Pool, *Technologies of Freedom* 169 & n.34 (1983) (citing Cabinet Committee on Cable Communications, *Cable—Report to the President* (1974)). *See also* Robert A. Kreiss, *Deregulation of Cable Television and the Problem of Access Under The First Amendment*, 54 S. Cal. L. Rev. 1001 (1981) (arguing that cable's monopoly status combined with its abundance of channels should give rise to affirmative duty to provide access to unaffiliated programmers).

international satellite system that would have rendered its satellites fully subject to traditional common carrier obligations.[68]

In the main, however, the regulatory regimes that developed around cable television "reflected the traditional view that the one-way delivery of television programs, movies, and sporting events is not a traditional common carrier activity and should not be regulated as such."[69] Accordingly, while "[t]he Commission has long had rules that require cable operators to reserve system capacity for programming produced by certain content providers unaffiliated with the cable operator,"[70] these rules have not required cable operators to interconnect their systems with those of other cable operators. Nor (except in the limited context of "leased access") have cable operators been required to provide any general right of channel access to unaffiliated content providers on a nondiscriminatory basis.[71] Analogously, until the advent of "direct access" in 1999,

---

[68] *See, e.g.*, S. 2650, 87th Cong. (1962) (proposing that the satellite system be constructed by a consortium of existing U.S. telecommunications common carriers); S. 2890, 87th Cong. (1962) (proposing that the government build, finance, and operate the proposed satellite system, and provide nondiscriminatory access to all carriers).

[69] *In re Inquiry Concerning High-Speed Access to the Internet Over Cable and Other Facilities, Declaratory Ruling & NPRM*, FCC 02-77, GEN Docket No. 00-185, 2002 WL 407567, ¶ 61 & n.232 (rel. Mar. 15, 2002). This "traditional view" was codified in the Cable Act of 1984. *See* 47 U.S.C. § 541(c) ("Any cable system shall not be subject to regulation as a common carrier or utility by reason of providing any cable service.").

[70] Howard Shelanski, *Regulating at the Technological Edge: New Challenges For the FCC*, 2000 L. Rev. Mich. St. U. Det. C.L. 3, 7 (2000). These rules include: the "must-carry" rules which require cable carriage of the broadcast signals of most local over-the-air broadcast channels, 47 U.S.C. §§ 534-35; the "PEG" rules which require carriage of a certain amount of noncommercial educational programming provided by the local franchising authority or its assignee, 47 U.S.C. § 531; and the "leased access" rules which reserve certain cable channels for hourly commercial lease by unaffiliated third parties, 47 U.S.C. § 532. The "must-carry" rules survived a First Amendment challenge in *Turner Broadcasting System v. FCC*, 520 U.S. 180 (1997). The "PEG" and "leased access" rules were substantially sustained in *Denver Area Educ. Telecommunications Consortium v. FCC*, 518 U.S. 727 (1996).

[71] In *AT&T Corp. v. City of Portland*, 216 F.3d 871, 876 (9th Cir. 2000), the Ninth Circuit suggested in *dicta* that because cable modem services are not "cable services" under the Communications Act, it "would lead to absurd results, inconsistent with the statutory structure" to construe the "leased access" provisions of 47 U.S.C. § 532 as vesting unaffiliated ISPs with a right to lease cable channel capacity in order to offer high-speed Internet service. Subsequently, the FCC's affirmed the Ninth Circuit's view that cable modem services are not "cable services" under the Act. *In re Inquiry Concerning High-Speed Access to the Internet Over Cable and Other Facilities, Declaratory Ruling & NPRM*, FCC 02-77, GEN Docket No. 00-185, 2002 WL 407567,

(continued . . .)

INTELSAT was not required to provide unbundled raw satellite capacity to unaffiliated international telecommunications carriers on a nondiscriminatory basis. Instead, INTELSAT's U.S. affiliate COMSAT enjoyed an exclusive right to furnish INTELSAT transmission capacity to carriers and users located in the United States, where such capacity was often bundled with other services provided by COMSAT.[72]

In lieu of relying on unbundling or interconnection requirements, regulators sought to control the monopoly power of both COMSAT and cable operators primarily through direct regulation of rates and services.[73] Thus, under the Satellite Act, the FCC exercised direct regulatory authority over the rates charged by COMSAT for INTELSAT transmission capacity,[74] and over the selection and the quality of COMSAT's service offerings.[75] Similarly, except during the late 1980s (when federal law preempted most local rate and service regulation), local cable franchising authorities sought to control the monopoly power of cable operators by directly regulating the rates and service offerings of cable franchisees.[76] In sum, for many years, there were substantial parallels between COMSAT and local cable operators with respect to both the nature of their respective "natural monopolies," and the regulatory responses thereto.

---

¶ 60 (rel. Mar. 15, 2002) . In so doing, the FCC appears to have eliminated any lingering possibility that the cable "leased access" rules might some day serve as a vehicle for implementing "cable open access." *Cf., e.g.*, Comments of Consumer and ISP Representatives at 3, 6-10; Comments of Consumers Union *et al.* at 22, *both filed in* GEN Docket No. 00-185 (filed Dec. 1, 2000) (each arguing for implementation of "cable open access" under the cable "leased access" rules).

[72]    *See* pages __, *supra*; *see also, e.g.*, Alexandra M. Field, Note, *INTELSAT At A Crossroads*, 25 Law & Pol'y Int'l Bus. 1335, 1341 (1994) (the Satellite Act "gave COMSAT a host of powers, the most important of which was *monopoly status in the provision of the U.S. satellite links*.") (emphasis added).

[73]    These regulatory tools were also included in the arsenal available to regulators of common carrier LECs. *See* 47 U.S.C. §§ 201, 203-05 (imposing rate regulation on telecommunications common carriers); *id.* §§ 208, 214 (imposing quality-of-service regulation on telecommunications common carriers). Unlike international satellite or cable operators, however, common carrier LECs were subject to access and unbundling requirements in addition to rate and quality-of-service regulation. *See* note __, *supra*.

[74]    *See* 47 U.S.C. § 741(a) (deeming COMSAT subject to common carrier rate regulation in its provision of INTELSAT satellite capacity and earth station services); 47 U.S.C. § 721(c)(2) (1994) (authorizing the FCC to regulate the "charges, classifications, practices, regulations, and other terms and conditions and [to] regulate the manner in which available facilities of the [satellite] system and [earth] stations are allocated among such users thereof") (repealed in 2000); *In re COMSAT Corp. Reclassification as a Non-Dominant Carrier*, 13 FCC Rcd. 14083, ¶ 8 (1998) ("*COMSAT Non-Dominant Order*") (noting that COMSAT's rates are subject to common carrier regulation under the Satellite Act), *modified on recon.*, 14 FCC Rcd. 3065 (1999). In 1998, COMSAT was substantially reclassified as a nondominant common carrier, and relieved from rate regulation on the large majority of its U.S.-international routes. *See id.* ¶¶ 2-3.

[75]    *Id.*; *see also* 47 U.S.C. § 721(c)(3)-(11) (1994)

### III. The Rise of Competition in the Cable Television and International Satellite Telecommunications Industries.

Despite the economic theory of "natural monopoly" and the legal and regulatory regimes that were constructed in its image, it turned out that—as with the telephone industry—putative competitors *did* wish to enter both the cable television and international satellite telecommunications industries, and to compete against the entrenched incumbent "natural monopolies."[77] Slowly, government responded to these competitors' pleas. During

---

[76] *See* House Report on Cable Communications Policy Act of 1984, H.R. Rep. No. 98-934, at 19 (1984), *reprinted in* 1984 U.S.C.C.A.N. 4655, 4656 ("Cable television has been regulated at the local government level through the franchise process. A municipal franchise granted to a cable operator has commonly specified . . . the service to be provided, and the rate which may be charged for those services." The 1984 Act substantially preempted such local rate and service regulation. *See* Pub. L. No. 98-549, 98 Stat. 2779 (1984); see also H.R. Rep. No. 98-934, at 24-25 (1984), *reprinted in* 1984 U.S.C.C.A.N. 4655, 4661-62 (asserting that some municipal franchising authorities had imposed unrealistic service requirements without permitting franchisees to price their services properly). In 1992, the Cable Television Consumer Protection and Competition Act substantially reinstated local government regulation of cable rates. *See* Pub. L. No. 102-385, 106 Stat. 1460 (1992); *see also Time Warner v. FCC*, 56 F.3d 151 (D.C. Cir. 1995) (sustaining FCC rules implementing the 1992 Act). Subsequently, however, the Telecommunications Act of 1996 heralded a partial return to deregulation by limiting the regulatory authority of local franchising authorities *only* to encompass rates charged for "basic tier" service, and not the rates charged for "cable programming" service (i.e. CNN, MTV, ESPN, etc.). *See* 47 U.S.C. § 543.

For critical discussions of the history of local regulation of retail cable TV rates, *see Time Warner Entertainment Co. v. FCC*, 56 F.3d 151, 161-205 (D.C. Cir. 1995). For critical discussions of the economic theory underlying local regulation of retail cable TV rates, *see* Donald J. Boudreaux & Robert B. Ekelund, Jr., *The Cable Television Consumer Protection and Competition Act of 1992: The Triumph of Private over Public Interest*, 44 Ala. L. Rev. 355, 362-66 (1993); Richard A. Posner, *The Appropriate Scope of Regulation in the Cable Television Industry*, 3 Bell J. Econ. & Mgmt. Sci. 98 (1972).

[77] *See* Glen O. Robinson, *The New Video Competition: Dances With Regulators*, 97 Colum. L. Rev. 1016, 1044 n.87 (1997) ("Today, the idea that cable should be treated as a natural monopoly is so unpopular not even the cable industry openly voices it. The basis for that unpopularity partly reflects skepticism about the degree of scale economies.

(continued . . .)

the 1980s, following the consent decree that broke up AT&T's telephone monopoly and introduced a measure of competition into long-distance telephony,[78] the U.S. government reversed its prior course and sought to permit and, indeed, encourage similar competition in the cable and satellite industries.

### A. The Introduction of Competition Against Incumbent Cable Television System Operators.

The first branch of government to act in favor of introducing facilities-based competition to the cable television industry was the judicial branch. In a series of cases beginning in the early 1980s, federal courts began to conceptualize local ordinances granting monopoly cable franchises as prior restraints against the "speech" of putative cable competitors, potentially subject to First Amendment limitation.[79] The Supreme Court tentatively adopted this view in *City of Los Angeles v. Preferred Communications, Inc.*,[80] in which it held that the First Amendment principle of the marketplace of ideas prohibits a municipality with excess utility pole capacity from awarding a monopoly cable franchise. Congress later followed the courts' lead, by providing in the 1992 Cable Act that "a franchising authority may not grant an exclusive franchise and may not unreasonably refuse to award an additional competitive franchise."[81]

Since the origin of intramodal cable competition in the 1980s, deployment of competing (or "overbuilt") systems has been slow.[82] Of the 33,000 cable community units

---

Even more, however, it reflects revised thinking about the policy implications: modern policy analysts do not see that large economies of scale necessarily foreclose the possibility or the desirability of active competition.") (citing Thomas W. Hazlett, *Private Monopoly and the Public Interest: An Economic Analysis of the Cable Television Franchise*, 134 U. Pa. L. Rev. 1335 (1986)). *Cf.* Stuart Minor Benjamin, Douglas Gary Lichtman, & Howard A. Shelanski, *Telecommunications Law and Policy* 379 (2001) (querying "why is it that often the legal response to natural monopoly is to restrict entry?" and whether "the entrance of a competitor into a market mean[s] that the market is not properly characterized as a natural monopoly?").

[78]  *United States v. AT&T*, 552 F. Supp. 131 (D.D.C. 1982), *aff'd sub nom.*, *Maryland v. United States*, 460 U.S. 1001 (1983).

[79]  *See, e.g., Community Communications Corp. v. City of Boulder*, 660 F.2d 1370, 1376-77 (10th Cir. 1981), *cert. dismissed*, 456 U.S. 1001(1982); *Omega Satellite Products v. City of Indianapolis*, 694 F.2d 119, 127 (7th Cir.1982); *Tele-Communications of Key West, Inc. v. United States*, 757 F.2d 1330, 1336-37 (D.C. Cir. 1985).

[80]  476 U.S. 488, 494-95 (1986).

[81]  Cable Act of 1992 § 621(a)(1), *codified at* 47 U.S.C. § 541(a)(1).

[82]  Cable "overbuilding" is defined as occurring "when two or more wireline cable
<space style="white-space: pre">                                              </space>(continued . . .)

<space style="white-space: pre">                                   </space>22

nationwide, "[a]s of year-end 1999, competing franchises [had] been awarded for service to [only] 369 communities in 34 states."[83] As of July 2000, the FCC had certified that overbuilt cable systems were actually operating and providing "effective competition" to the incumbent providers in 330 of these 369 individual communities.[84] In January, 2002, however, the FCC declined to provide an updated count of overbuilt communities, while suggesting that many overbuilders "are facing difficulties in obtaining capital . . . [and therefore] have scaled back plans, reduced capital expenditures, reduced staffs, or shut down operations altogether."[85]

On the other hand, cable television since the 1980s has faced increasing intermodal competition from other, newer multichannel video programming distributors

---

systems directly compete for subscribers in a local delivery market." *Notice of Inquiry Concerning Deployment of Advanced Telecommunications Capability to All Americans in a Reasonable and Timely Fashion*, 15 FCC Rcd. 16641, ¶ 45 n.66 (2000) (citation omitted).

[83] *In re Annual Assessment of Status of Competition in Market for Delivery of Video Programming*, *Seventh Annual Report*, 16 FCC Rcd. 6005, 22 Comm. Reg. (P&F) 1414, ¶ 37 & n.110 (2001). Because some of the nation's most densely populated cable communities number among the 369 in which competitive systems have been authorized, however, franchised cable overbuilders currently have the potential to serve more than 18.5 million [of America's 105.5 million television] homes. *Id.* By way of comparison, as of June 2001, 69 million homes subscribed to cable television service provided by incumbent franchisees, and another 1.5 million subscribed to non-franchised "SMATV" service, which uses substantially the same technology used by franchised cable TV systems, but serves apartment buildings without using public rights-of-way. *In re Annual Assessment of Status of Competition in Market for Delivery of Video Programming*, *Eighth Annual Report*, 17 FCC Rcd. 1244, ¶ 7, 13 (2002). Only twenty percent of American households (*i.e.*, about 20.7 million households) currently watch television via free, over-the-air terrestrial broadcast signals. *Id.* ¶ 79.

[84] *Id.* ¶ 37; *see also* 47 U.S.C. § 543(l)(1) (defining four possible alternative bases for determining that an incumbent cable TV system is now subject to "effective competition," and hence relieved from rate regulation).

[85] *In re Annual Assessment of Status of Competition in Market for Delivery of Video Programming*, *Eighth Annual Report*, 17 FCC Rcd. 1244, ¶ 108 (2002). The Commission in the *Eighth Annual Report* did announced that "[o]f the 33,000 cable community units nationwide, 419, or approximately one percent have been certified by the Commission as having effective competition as a result of consumers having a choice of more than one MVPD." *Id.* ¶ 120. This new figure of 419, however, includes not just communities that have been overbuilt by a competing wireline system, but also those in which DBS satellite has gained a market share sufficient to provide effective intermodal competition to the incumbent cable operator.

("MVPDs"), which can provide many of the same (or substitutable) services.[86] Most prominent among these new competing MVPDs are satellite dish providers.[87] Cable is also subject to limited (but potentially increasing) competition from "Open Video Systems" that can be provided by Local Exchange Carriers ("LECs") or electric utility companies through modified telephone or electrical lines.[88] At the low end, cable continued to face (price) competition from conventional broadcast television, from home video sales and rentals, and from terrestrial "wireless cable."[89] Accordingly, while the FCC recently concluded that "[c]able television still is the dominant technology for the delivery of video programming to consumers in the MVPD marketplace," it also noted that cable's market share continues to decline.[90]

### B. The Introduction of Competition Against INTELSAT.

Although Congress initially contemplated the INTELSAT system to be a "natural monopoly" with respect to international satellite telecommunications, the FCC as early as 1966 proposed to authorize the construction and operation of separate, non-INTELSAT satellite systems to provide *domestic* satellite service, both to augment the long-haul terrestrial facilities of existing telecommunications carriers for point-to-point switched transmission services, and also to connect off-shore distant domestic points (*i.e.*, Alaska,

---

[86] As of June 2001, a total of 19.3 million American homes were subscribing to non-cable multichannel video programming distributors ("MVPDs"), most prominently Direct Broadcast Satellite (DBS) service. *In re Annual Assessment of Status of Competition in Market for Delivery of Video Programming*, *Eighth Annual Report*, 17 FCC Rcd. 1244, ¶¶ 7, 13 (2002).

[87] *Id.* ¶ 13. The FCC characterizes all residential satellite dish service collectively as "Direct-to-Home" ("DTH") satellite service. The DTH category consists of two subcategories: the popular "DBS" service (which generally uses a twelve-inch satellite dish), and the older but less popular "HSD" service (which requires a much larger dish). As of June 2001, sixteen million American households were subscribing to DBS service, and another 1 million to HSD. *Id.* Collectively, the two DTH satellite services are capable of providing service to virtually every U.S. home that cable can serve, as well as some homes not yet passed by cable.

[88] *Id.*; *see also* 47 U.S.C. § 543(l)(1)(D). For a history of the FCC's policy transition from prohibiting to encouraging the deployment of OVS, *see* Glen O. Robinson, *The New Video Competition: Dances With Regulators*, 97 Colum. L. Rev. 1016 (1997).

[89] *In re Annual Assessment of Status of Competition in Market for Delivery of Video Programming*, *Eighth Annual Report*, 17 FCC Rcd. 1244, ¶¶ 13, 79 (2002).

[90] *Id.* ¶ 2.

Hawaii, Puerto Rico) to the contiguous states.[91] In so proposing, the FCC also contemplated that non-INTELSAT domestic satellites might some day be used to provide point-to-multipoint services, such as broadcast program transmission.[92] In 1970, the FCC acted on its earlier proposals by adopting an "open skies" policy under which qualified private applicants could be authorized to construct and operate domestic satellite systems capable of competing against both the domestic terrestrial carriers and INTELSAT in the long-haul domestic telecommunications market.[93] The "open skies" policy paved the way for the development of a domestic satellite communications industry.

Even while crafting an "open skies" policy for domestic satellite communications, however, the FCC's *DOMSAT I* and *DOMSAT II Orders* protected INTELSAT against the advent of intramodal competition in the market for U.S.-international satellite telecommunications capacity. In part, the FCC's decision to preserve INTELSAT's monopoly was influenced by the terms of the INTELSAT Agreement, which permitted INTELSAT itself to veto the development of any separate international satellite systems that would cause "economic harm" to INTELSAT.[94] However, as international telecommunications traffic expanded rapidly during the 1970s and 1980s, INTELSAT did consent to the development of several competing "separate systems" that provided competitive regional international service in different parts of the world.[95]

---

[91] *See Notice of Inquiry*, 31 Fed. Red. 3507 (Mar. 2, 1966); *Supplemental Notice of Inquiry*, 31 Fed. Reg. 13763 (Oct. 20, 1966).

[92] *Id.*; *see also Establishment of Domestic Communications-Satellite Facilities By Non-Governmental Entities, Second Report and Order*, 35 FCC 2d 844, ¶ 4 (1972) ("*DOMSAT II Order*"), *recon. denied*, 38 FCC 2d 665 (1972), *aff'd sub nom*, *Network Project v. FCC*, 511 F.2d 786 (D.C. Cir. 1975).

[93] *See Establishment of Domestic Communications-Satellite Facilities By Non-Governmental Entities, First Report and Order*, 22 FCC 2d 86, 93 (1970) ("*DOMSAT I Order*").

[94] *Agreement Relating to the International Telecommunications Satellite Organization "INTELSAT,"* Art. XIV(d), done Aug. 20, 1971, 23 U.S.T. 3813 ("INTELSAT Agreement"). The "economic harm" requirement was justified as a means of protecting INTELSAT against "cream-skimming" in order to safeguard INTELSAT's ability to serve every country on earth, regardless of cost, on non-discriminatory terms and conditions. *See* Charles H. Kennedy & M. Veronica Pastor, *An Introduction To International Telecommunications Law* 79-80 (1996).

[95] *See* Charles H. Kennedy & M. Veronica Pastor, *An Introduction To International Telecommunications Law* 80 (1996) ("The first separate system to receive approval was [Western Europe's] EUTELSAT in 1979, which was soon followed by [Southeast Asia's] PALAPA and [the Middle East's] ARABSAT.").

On November 28, 1984, President Reagan "determine[d] that separate international communications satellite systems [were] required in the national interest."[96] Accordingly, both the Secretary of State and the Secretary of Commerce were jointly directed "to inform the Federal Communications Commission of criteria necessary to ensure the United States meets its international obligations [under the INTELSAT Agreement] and to further its telecommunications and foreign policy interests" by establishing separate satellite systems to compete against the INTELSAT system.[97] In 1985, the FCC responded to the President's directive by authorizing, for the first time, the licensing of separate international satellite systems.[98] In 1988, the Connecticut-based PanAmSat Corp. launched the PAS-1 Atlantic Ocean Region satellite, the first U.S. private-sector satellite to provide international satellite services.[99] By 1999, more than 200 commercial geosynchronous satellites were orbiting the earth, of which approximately 73 served the United States.[100] Of these, only 17 satellites belonged to INTELSAT, of which just 13

---

[96] Presidential Determination No. 85-2, 49 Fed. Reg. 46987, 1984 WL 88118 (Nov. 28, 1984). For discussions of the deliberations that led to this Presidential determination, see Bert W. Rein & Carl R. Frank, *The Legal Commitment of the United States to the INTELSAT System*, 14 N.C. J. Int'l L. & Com. Reg. 219, 225-27 (1989); Richard R. Colino, *A Chronicle of Policy and Procedure: The Formulation of the Reagan Administration Policy on International Satellite Telecommunications*, 13 J. Space L. 103 (1985); Richard R. Colino, *The Possible Introduction of Separate Satellite Systems: International Satellite Communications at the Crossroad*, 24 Colum. J. Transnat'l L. 13 (1985).

[97] Presidential Determination No. 85-2, 49 Fed. Reg. 46987, 1984 WL 88118 (Nov. 28, 1984).

[98] *See Establishment of Satellite Systems Providing International Communications*, 101 FCC 2d 1046 (1985) ("*Separate Systems Order*"), *modified on recon.*, 61 Rad. Reg. 2d (P&F) 649 (1986), *further recon. denied*, 1 FCC Rcd 439 (1986). Until 1996, these separate international systems were restricted from interconnecting directly with the Public Switched Telephone Network (PSTN). *Compare id.* (imposing PSTN restriction) *with Permissible Services of U.S.-Licensed International Communications Satellite Systems Separate From INTELSAT*, 7 FCC Rcd. 2313 (1992) (sunsetting the PSTN restriction effective January 1, 1997). *See also In re COMSAT Corp. Reclassification as a Non-Dominant Carrier*, 13 FCC Rcd. 14083, ¶ 59 (1998) (noting that the PTSN restrictions did, in fact, sunset on January 1, 1997), *modified on recon.*, 14 FCC Rcd. 3065 (1999).

[99] *See* PanAmSat History Web Page, <http://www.panamsat.com/comp/history.htm>. In 1996, PanAmSat was acquired by the Hughes Electronics Corporation, which is itself 81%-owned by the General Motors Corporation. *Id.* Today, Hughes/PanAmSat operates a fleet of 21 geosynchronous satellites—the largest in the world, and two more than INTELSAT. *Id.*

[100] *See Phillips Satellite Industry Directory*, at 17-234, 279-413 (21st ed. 1999)

served the United States.[101] Accordingly, in 1998, the FCC found that INTELSAT was subject to substantial intramodal competition on most of its U.S.-international routes.[102]

At the same time, INTELSAT also began to face substantial intermodal competition in the market for international communications transmission capacity. During the 1990s, the world witnessed a proliferation of high-capacity transoceanic submarine fiber optic cables that are capable of delivering many of the same services that satellites can deliver, often at lower cost.[103] In fact, since the early 1990s, fiber-optic cable systems have carried more traffic for U.S.-international switched voice and private line than satellite systems have.[104] "Excluding traffic carried to Mexico and Canada over terrestrial networks, markets COMSAT does not serve, fiber-optic cable systems carried three times as much switched voice traffic and six times as much private line traffic than satellite networks in 1996."[105]

---

(setting forth complete information about each of these satellites and their operators).

[101] *See In re Availability of INTELSAT Space Segment Capacity To Users and Service Providers Seeking To Access INTELSAT Directly*, 15 FCC Rcd. 19160, ¶¶ 2, 5 (2000).

[102] *See COMSAT Non-Dominant Order*, 13 FCC Rcd. 14083, ¶ 59 (1998) ("A number of satellite systems are significant competitors [to INTELSAT] for the full-time video and occasional-use video service markets. These other satellite systems include U.S.-based systems, such as PanAmSat and Orion, as well as [foreign-licensed] regional satellite service providers.").

[103] *See, e.g. In re COMSAT Corp. Reclassification as a Non-Dominant Carrier*, 13 FCC Rcd. 14083, ¶¶ 11, 19, 32-39 (1998) ("*COMSAT Non-Dominant Order*"), *modified on recon.*, 14 FCC Rcd. 3065 (1999) (characterizing satellites and submarine cables as fungible commodities serving the markets for switched voice, private line, and video services, and noting that cables compete effectively against INTELSAT satellites on every major international telecommunications route to or from the United States); *see also Changes in International Satellite Policy: Hearing Before the Subcomm. on Communications of the Sen. Committee on Commerce, Science and Transportation*, 1999 WL 170205 (March 25, 1999) (Testimony of INTELSAT Director General Conny Kullman) (providing detailed information about various submarine cable and separate satellite systems).

[104] *COMSAT Non-Dominant Order*, 13 FCC Rcd. 14083, ¶ 56 (1998); *see also id.* ¶ 76 ("Intermodal competition leads us to believe that fiber-optic cables represent a substitute for satellites in the transmission of switched voice service.").

[105] *Id.*

For these and other[106] reasons, the FCC in 1998 reclassified COMSAT/ INTELSAT as a non-dominant carrier on every major telecommunications route to or from the United States.[107] By so doing, the FCC found that, with respect to these routes, COMSAT lacked "market power"—*i.e.*, that it possessed neither "[c]ontrol of bottleneck facilities,"[108] nor the "ability to raise or maintain prices above costs, control prices, or exclude competition."[109] This reclassification occurred one year *before* the FCC was to implement "direct access" to INTELSAT.[110] Indeed, while COMSAT's customers and competitors urged the FCC "to permit other U.S. providers and users to obtain direct access to INTELSAT before non-dominant relief is granted,"[111] the FCC rejected such demands.[112] Accordingly, the FCC clearly considered INTELSAT to be "non-dominant" in the international telecommunications market at the time it imposed unbundling obligations on INTELSAT. Incumbent cable television system operators, in contrast, continue to this day to be "dominant" in their markets, but are not subject to analogous unbundling obligations.[113]

---

[106] The FCC in the *COMSAT Non-Dominant Order* also stated its expectation that INTELSAT would soon face further competition from several Low Earth Orbit satellite systems (LEOs), such as Iridium, that were then being deployed. In theory, LEOs can provide many of the same services that GEOs can provide, with the additional benefit of mobile use. In fact, however, Iridium's spectacular failure to attract customers resulted in the company's prompt $5 billion bankruptcy. *See* Nicole Harris, *Iridium to End Satellite Service, Liquidate Assets*, Mar. 20, 2000, Wall St. J., at B8, 2000 WL-WSJ 3022232. Other LEO systems, such as the $4.3 billion GlobalStar, have followed Iridium into bankruptcy. *See, e.g.*, Andy Pasztor, *Globalstar's Chapter 11 Filing Reflects Lack of Restructuring Plan, Customers*, Feb. 19, 2002, Wall St. J., at B6, 2002 WL-WSJ 3386238.

[107] *COMSAT Non-Dominant Order*, 13 FCC Rcd. 14083 (1998), *modified on recon.*, 14 FCC Rcd. 3065 (1999). At the same time, the FCC retained its classification of COMSAT as a dominant carrier on a number of minor "thin routes," that were not served by any cable or satellite provider other than INTELSAT. These routes primarily connected the United States with "developing nations located in Africa and Eastern Europe as well as low density, remotely located island nations, such as Mauritius and New Caledonia, that might not justify the cost of a cable connection." *Id.* ¶ 28; *see also id.* App. A (listing INTELSAT's "thin routes" for each product market). In a subsequent *Order*, the FCC imposed price-cap regulation on COMSAT's rates for transmission capacity over its remaining thin routes. *Policies and Rules for Alternative Incentive Based Regulation of Comsat Corp.*, 14 FCC Rcd. 3065 (1999).

[108] *COMSAT Non-Dominant Order*, 13 FCC Rcd. 14083, ¶ 9 (1998).

[109] *Id.* ¶ 66. As discussed above, these findings pertained only to COMSAT's major U.S. international routes, and not to the "thin routes" served by no other satellite or cable service provider. In 1999, "thin route" service accounted for less than 8 percent of COMSAT's INTELSAT-based revenues. *See Policies and Rules for Alternative Incentive Based Regulation of Comsat Corp.*, 14 FCC Rcd. 3065 (1999).

[110] Direct access was finally implemented in *Direct Access To The INTELSAT System, Report & Order*, 14 FCC Rcd 15703 (1999) ("*1999 Direct Access Order*"). *See* Subpart V.A, *infra*.

### IV. The Introduction of Residential High-Speed Internet Access.

In the late 1990s, large numbers of consumers began to seek access to the Internet directly from their homes, in order to use the World Wide Web.[114] Until 1996, most residential users could access the Internet only by dialing up an Internet Service Provider (ISP), such as America Online (AOL), via local telephone calls in which digital information was translated through analog voice-grade dial-up modems incapable of translating more than 56 kilobytes per second (kbps) of information.[115] Today, dial-up ISPs that provide

---

[111] *See COMSAT Non-Dominant Order*, 13 FCC Rcd. 14083, ¶ 17 & n.49 (1998).

[112] *See COMSAT Non-Dominant Order*, 13 FCC Rcd. 14083, ¶ 180 (1998) ("We grant Comsat's request for reclassification as a non-dominant common carrier with respect to its provision of INTELSAT services in the switched voice, private line, full-time video, and occasional-use video services to competitive markets. We also find Comsat non-dominant in the provision of earth station services"); *id.* ¶ 156 ("our action today does not require direct access to INTELSAT or a waiver by Comsat of its immunity"). *Accord 1999 Direct Access Order*, 14 FCC Rcd 15703, ¶ 12 (1999) (same).

[113] *See* page __, *supra* (discussing continuing dominance of cable TV in local MVPD markets); *accord In re Annual Assessment of Status of Competition in Market for Delivery of Video Programming, Eighth Annual Report*, 17 FCC Rcd. 1244, ¶ 2 (2002) (same). *See also* page __, *infra* (discussing continuing dominance of cable modem service in local residential high-speed Internet access markets); *accord In re Inquiry Concerning High-Speed Access to the Internet Over Cable and Other Facilities, Declaratory Ruling & NPRM*, FCC 02-77, GEN Docket No. 00-185, 2002 WL 407567, ¶ 9 (rel. Mar. 15, 2002) (same).

[114] Internet use in the United States has grown at a rate of 20 percent a year since 1998. U.S. Department of Commerce, National Telecommunications and Information Administration, *A Nation Online: How Americans Are Expanding Their Use of the Internet* 10 (Feb. 2002), available online at <http://www.ntia.doc.gov/ntiahome/dn/anationonline2.pdf>. As of September 2001, 53.9 million U.S. households (50.5%) had Internet connections. *Id.* at 3.

[115] Although some local exchange carriers companies (LECs) offered their own proprietary ISP services, these LECs were required, as common carriers, to provide transmission services to competing ISPs, and thereby effectively to unbundle their own proprietary ISP service from their telephone services. *See Notice of Inquiry Concerning*

(continued . . .)

low-speed "narrowband" connections to the Internet continue to provide the lion's share of residential Internet access throughout the United States.[116]

Despite the dominant position of "narrowband" ISP service in the current residential ISP market, however, an increasing number of users consider "narrowband" connections to be unsuitable for downloading the data-intensive audiovisual content and applications that have proliferated on the World Wide Web.[117]  Accordingly, new "broadband" services offering high-speed Internet connections to residential subscribers are rapidly gaining in popularity.[118]  At the end of 1999, only 1.8 million Americans were residential subscribers to high-speed Internet services.[119]  By the end of 2000, this number had risen to 6.8

---

*High-Speed Access to Internet*, 15 FCC Rcd. 19287, ¶ 6 (2000)  (noting that for dial-up Internet access, "'last mile' transmission capability is available independently of the choice of ISP").  Partly for this reason, the market for "narrowband" dial-up ISP service was quite competitive nearly from its inception.

[116]    As of August 2001, 80% of the nation's residential Internet users still obtained Internet access via low-speed "dial-up" service.  U.S. Department of Commerce, National Telecommunications and Information Administration, *A Nation Online: How Americans Are Expanding Their Use of the Internet* 2 (Feb. 2002), available online at <http://www.ntia.doc.gov/ntiahome/dn/anationonline2.pdf>.

[117]    *See In re Inquiry Concerning High-Speed Access to the Internet Over Cable and Other Facilities, Declaratory Ruling & NPRM*, FCC 02-77, GEN Docket No. 00-185, 2002 WL 407567, at ¶ 10 (rel. Mar. 15, 2002) (discussing limitations of "narrowband" access and advantages of high-speed access).

[118]    *See* U.S. Department of Commerce, National Telecommunications and Information Administration, *A Nation Online: How Americans Are Expanding Their Use of the Internet* 2 (Feb. 2002), *available online at* <http://www.ntia.doc.gov/ntiahome/dn/anationonline2.pdf> (noting that "[b]etween August 2000 and September 2001, residential use of high-speed, broadband service doubled—from about 5 to 11 percent of all individuals, and from 11 to 20 percent of Internet users.").  *See also Notice of Inquiry Concerning High-Speed Access to Internet*, 15 FCC Rcd. 19287, ¶¶ 6-7 (2000) (describing transition from dial-up access to high-speed access).  As a matter of terminology, the FCC defines "high-speed" transmission services to include "those services and facilities with a transmission speed of 200 kbps in at least one direction."  *In re Inquiry Concerning the Deployment of Advanced Telecommunications Capability to All Americans in a Reasonable and Timely Fashion, Second Report*, 15 FCC Rcd. 20913, ¶ 11 (2000). "Advanced telecommunications capability and advanced services," in contrast, form a subset of the larger "high-speed" category including only those services and facilities "capable of 200 kbps or greater transmission in *both* directions."  *Id.* (emphasis added).

[119]    *In re Inquiry Concerning the Deployment of Advanced Telecommunications Capability to All Americans in a Reasonable and Timely Fashion, Second Report*, 15 FCC

million.[120]  "[A]s of June 30, 2001, about 7.8 million households subscribed to high-speed services," which by then were available to approximately 75-80% of all the homes in the United States via DSL or cable modem service.[121]  By the end of 2002, industry analysts project that "total U.S. broadband subscribers will rise . . . to over 19 million."[122]  "By 2008, that number is predicted to reach 78 million.[123]

At present, high-speed Internet connections are delivered to residential users via several competing technological platforms.  In most U.S. locations, incumbent cable operators became the first entities to deploy high-speed Internet access to residential users.  Cable operators were able to do this by upgrading their existing HFC cable facilities to offer two-way high-speed data transmission services, with Internet access made possible through cable modem technologies.[124]  Cable Internet connections are very fast, often

---

Rcd. 20913,  ¶ 8(2) (2000).

[120]     Ernie Bergstrom, Mike Paxton, & Michelle Abraham, *The Broadband Marathon: Access Technologies Jockey for Subscribers* (Cahners In-Stat Group Rep. No. MB01-04MI) (June 22, 2001), *abstracted online at* <http://www.adsl.com/latestnews/analyst_corner.html>.

[121]     *In re Inquiry Concerning High-Speed Access to the Internet Over Cable and Other Facilities, Declaratory Ruling & NPRM*, FCC 02-77, GEN Docket No. 00-185, 2002 WL 407567, at ¶ 9 & n.24 (rel. Mar. 15, 2002) (citing *Deployment of Advanced Telecommunications Capability to All Americans in Reasonable and Timely Fashion, Possible Steps to Accelerate Such Deployment Pursuant to Section 706 of Telecommunications Act of 1996, Third Report*, FCC 02-33, CC Docket 98-146, 2002 WL 186930, at ¶ 7 (rel. Feb 6, 2002)).  In addition, many of the 20-25% of U.S. homes that cannot obtain cable modem or DSL service can obtain high-speed Internet access via satellite.  *See* note __, *infra*.

[122]     Ernie Bergstrom, Mike Paxton, & Michelle Abraham, *The Broadband Marathon: Access Technologies Jockey for Subscribers* (Cahners In-Stat Group Rep. No. MB01-04MI) (June 22, 2001), *abstracted online at* <http://www.adsl.com/latestnews/analyst_corner.html>.

[123]     Deborah A. Lathen, *Broadband Today, FCC Staff Report on Industry Monitoring Sessions* 9 (1999), *available online at* <http://www.fcc.gov/Bureaus/Cable/Reports/broadbandtoday.pdf> (citing John Schwartz, *How Much Room in the Fat Pipe?*, Sept. 19, 1999, Wash. Post, at H01)).

[124]     *See In re Inquiry Concerning the Deployment of Advanced Telecommunications Capability to All Americans in a Reasonable and Timely Fashion, Second Report*, 15 FCC Rcd. 20913,  ¶ 29 (2000) ("Cable modem technologies rely on the same basic network architecture used for many years to provide multichannel video service, but with upgrades and enhancements to support advanced services."); *see also id.* ¶ 31 (explaining the upgrades needed).

providing downstream transmission speeds of up to 1.5 megabytes per second (Mbps) to residential users.[125] Moreover, in launching high-speed Internet service, cable operators realized substantial savings by utilizing the same cable that had already been laid to provide cable TV service to reach the homes of high-speed Internet subscribers.[126]

Unlike incumbent LECs,[127] "cable operators currently are not legally prohibited from having an exclusive relationship with one particular ISP," including an affiliated or wholly-owned ISP.[128] Accordingly, most cable operators bundled three distinct services into a single package collectively marketed as "cable modem service": (1) high-speed cable transmission of data between the user's home and the cable headend; (2) access to the Internet through proprietary servers and routers located at the cable headend, plus management of the use of the cable network for data delivery services; and (3) browsing and e-mail functionalities, some online content (usually including a search engine), and user support services similar in nature to those offered by other ISPs.[129] In short, residential cable modem service tends to combine transmission services and ISP services into a single bundled package.[130]

---

[125] *Id.* ¶ 33.

[126] *In re Inquiry Concerning the Deployment of Advanced Telecommunications Capability to All Americans in a Reasonable and Timely Fashion, Second Report*, 15 FCC Rcd. 20913, ¶ 31 (2000) ("Once an HFC network is upgraded, new services are available to all homes passed by the upgraded infrastructure. This contrasts with DSL technologies, where variations in legacy outside plant conditions can limit access to certain end-users even in upgraded areas, and with wireless technologies where line-of-sight requirements may be a factor.").

[127] *See In re Matters of Deployment of Wireline Services Offering Advanced Telecommunications Capability, Third and Fourth Report and Order*, 14 FCC Rcd. 20912, ¶ 4 (1999) ("*Line Sharing Order*") (amending unbundling rules to require incumbent LECs to provide unbundled access to the high-frequency portion of the local loop, in order to enable unaffiliated competitors to provide DSL service through the incumbent LEC's telephone lines), *recon. denied in pertinent part*, 16 FCC Rcd. 2101 (2001), *appeal pending sub nom., United States Telecom Association v. FCC*, No. 00-1012 (D.C. Cir. filed Jan. 18, 2000); *see also Appropriate Framework for Broadband Access to the Internet Over Wireline Facilities*, *Universal Service Obligations of Broadband Providers*, *Notice of Proposed Rulemaking*, FCC 02-42, CC Docket No. 02-33, 2002 WL 252714 (rel. Feb. 15, 2002) (initiating proceeding to consider whether DSL unbundling requirements should be relaxed to attain regulatory parity with cable modem service).

[128] *Notice of Inquiry Concerning High-Speed Access to Internet*, 15 FCC Rcd. 19287, ¶ 29 (2000). The FCC did, however, recently initiate a new rulemaking proceeding to "consider whether it is necessary or appropriate at this time to require that cable operators provide unaffiliated ISPs with the right to access cable modem service customers directly." *In re Inquiry Concerning High-Speed Access to the Internet Over Cable and Other Facilities, Declaratory Ruling & NPRM*, FCC 02-77, GEN Docket No. 00-185, 2002 WL 407567, ¶ 72 (rel. Mar. 15, 2002).

Because very few communities currently enjoy two or more overbuilt HFC cable systems,[131] very few incumbent cable operators face intramodal competition in the provision of high-speed Internet service. Intermodal competition, however, arose soon after the deployment of cable modem service, and is continuing to spread quickly. As of August 2001, cable modems continued to serve 68% of the market for residential high-speed Internet access, but this market share appears still to be dropping.[132]

Chief among cable's intermodal competitors in the residential high-speed Internet access market is asymmetric digital subscriber line service ("ADSL" or "DSL"), which is capable of delivering a 768 kbps download and a 384 kbps upload through reconfigured "twisted pair" copper wires of the existing telephone system.[133] Although DSL service has been widely deployed by LECs in U.S. metropolitan areas, the service is subject to certain technical limitations that currently prevent it from being deployed to some potential residential end-users.[134] Nonetheless, DSL has already captured 29% of the U.S. residential high-speed Internet access market.[135] As noted above, unlike cable modem providers, most incumbent LECs are required by FCC Rules to unbundle their transmission services from their ISP services.[136]

---

[129] *Notice of Inquiry Concerning High-Speed Access to Internet*, 15 FCC Rcd. 19287, ¶ 10 (2000). *See also In re Inquiry Concerning High-Speed Access to the Internet Over Cable and Other Facilities, Declaratory Ruling & NPRM*, FCC 02-77, GEN Docket No. 00-185, 2002 WL 407567, ¶ 11 (rel. Mar. 15, 2002) ("Cable operators often include in their cable modem service offerings all of the services typically provided by Internet access providers, so that subscribers usually do not need to contract separately with another Internet access provider to obtain discrete services or applications, such as an e-mail account or connectivity to the Internet, including access to the World Wide Web") (footnote omitted). For present purposes, the FCC defines "cable modem service" as "a service that uses cable system facilities to provide residential subscribers with high-speed Internet access, as well as many applications or functions that can be used with high-speed Internet access." *In re Inquiry Concerning High-Speed Access to the Internet Over Cable and Other Facilities, Declaratory Ruling & NPRM*, FCC 02-77, GEN Docket No. 00-185, 2002 WL 407567, ¶ 31 (rel. Mar. 15, 2002).

[130] *See National Cable & Telecommunications Ass'n, Inc. v. Gulf Power Co.*, 122 S. Ct. 782, 796-97 n.4 (2002) (Thomas, J., concurring in part and dissenting in part) ("Residential high-speed Internet access typically requires two separate steps. The first is transmission from a customer's home to an Internet service provider's (ISP's) point of presence. This service is generally provided by a cable or phone company over wires attached to poles, ducts, conduits, and rights of way. The second is a service delivered by an ISP to provide the connection between its point of presence and the Internet.").

[131] *See* note __, *supra*.

[132] *In re Inquiry Concerning High-Speed Access to the Internet Over Cable and Other Facilities, Declaratory Ruling & NPRM*, FCC 02-77, GEN Docket No. 00-185, 2002 WL 407567, ¶ 9 (rel. Mar. 15, 2002); *accord* U.S. Department of Commerce, National Telecommunications and Information Administration, *A Nation Online: How Americans Are Expanding Their Use of the Internet* 35 (Feb. 2002), available online at <http://www.ntia.doc.gov/ntiahome/dn/anationonline2.pdf>.

In addition to cable and DSL (which collectively now control 97% of the residential high-speed Internet access market), several other emerging technologies also can deliver high-speed Internet data transmissions to residential users. For example, several terrestrial "fixed wireless" technologies currently enable two-way digital signals to be transmitted through the wireless spectrum between a local transmitting tower and a special antenna that must be installed on the user's rooftop.[137] Like cable system operators and unlike

---

[133] *See, e.g.*, Cincinnati Bell Zoomtown Web Page, <http://company.zoomtown.com/zt_fs.html>.

[134] *See In re Inquiry Concerning the Deployment of Advanced Telecommunications Capability to All Americans in a Reasonable and Timely Fashion, Second Report*, 15 FCC Rcd. 20913, ¶¶ 38-40 (2000) (explaining technical reasons why ADSL cannot serve many American homes).

[135] *In re Inquiry Concerning High-Speed Access to the Internet Over Cable and Other Facilities, Declaratory Ruling & NPRM*, FCC 02-77, GEN Docket No. 00-185, 2002 WL 407567, ¶ 9 (rel. Mar. 15, 2002). The NTIA estimates DSL's share of the high-speed market to be 33%. U.S. Department of Commerce, National Telecommunications and Information Administration, *A Nation Online: How Americans Are Expanding Their Use of the Internet* 35 (Feb. 2002), available online at <http://www.ntia.doc.gov/ntiahome/dn/anationonline2.pdf>.

[136] *See In re Matters of Deployment of Wireline Services Offering Advanced Telecommunications Capability, Third and Fourth Report and Order*, 14 FCC Rcd. 20912, ¶ 4 (1999) ("*Line Sharing Order*"), discussed *supra* note __. In practice, this unbundling requirement enables a user whose DSL line is provided by her local telephone company (*e.g.*, Bell Atlantic or BellSouth) to choose to obtain access to the Internet through proprietary servers and routers of an unaffiliated ISP (*e.g.*, AOL or EarthLink) without also having to pay for the use of the telephone company's servers and routers.

[137] *See generally In re Inquiry Concerning the Deployment of Advanced Telecommunications Capability to All Americans in a Reasonable and Timely Fashion, Second Report*, 15 FCC Rcd. 20913, ¶¶ 42-55 (2000) (surveying various fixed wireless technologies). In Chicago, one such fixed wireless service currently provides typical downstream transmission speeds of 512 kbps to 1.5 Mbps, with a 256 kbps maximum

(continued . . .)

incumbent LECs, providers of fixed wireless services are not required by law to unbundle their transmission services from their ISP services.

High-speed data transmissions also can be delivered to residential users via satellite. At present, at least one geostationary satellite system, Hughes' DirecPC, can use the satellites of its DBS affiliate, DirecTV, to provide 400 kbps downstream transmission to any household in the continental United States that has an unobscured line-of-sight to the southern sky.[138] Moreover, newly developing technologies will use low-earth-orbit satellites (LEOs) to offer mobile, two-way communications, at even higher transmission speeds than existing technologies. The most ambitious such system, the Teledesic Network, is currently building a proposed 288-satellite system that promises to provide millions of simultaneous users with downlink transmission speeds up to 64 Mbps, and uplink speeds up to 2 Mbps.[139] Like cable system operators and terrestrial fixed wireless service providers, but unlike incumbent LECs, satellite providers are not required by law to unbundle their transmission services from their ISP services.

In sum, cable modem service emerged as the first entrant into the market for residential high-speed Internet access, and therefore briefly enjoyed *de facto* monopoly status in that market. Because regulators never conceived of residential high-speed Internet access as a "natural monopoly," however, cable modem service providers have not been shielded by law against competition, nor have they been subject to regulation of their rates or service offerings or to mandatory unbundling requirements.

---

upload speed. *See* Sprint Broadband Direct Web Page, <http://www.sprintbroadband.com/>.

[138] *See generally In re Inquiry Concerning the Deployment of Advanced Telecommunications Capability to All Americans in a Reasonable and Timely Fashion, Second Report*, 15 FCC Rcd. 20913, ¶¶ 56-59 (2000). Because DBS satellites are not equipped for two-way communications, upstream DirecPC transmissions must use conventional telephone dial-up connection. *Id.* ¶ 58; *see also* DirecPC Web Page, <http://www.direcpc.com/>. However, newer systems are now beginning to offer *two-way* high-speed Internet data transmission (and ISP) services entirely via geostationary satellites, without the need for a dial-up uplink. *See, e.g.*, StarBand Web Page, <http://www.starband.com/howitworks/index.htm>. INTELSAT, in contrast, while serving as an important supplier of Internet backbone to many ISPs, does not function as an ISP, nor does it use its geostationary satellites to provide any data transmission services directly to residential users. *See* Intelsat Web Page, <http://www.intelsat.int/services_internet.asp>.

[139] Teledesic Web Page, <http://www.teledesic.com/about/about.htm>; *see also* SkyBridge Web Page, <http://www.skybridgesatellite.com/l41_sys/index.htm> (describing another proposed LEO system that, beginning in 2005, promises to deliver two-way residential service with 5 Mbps downlink transmissions from a constellation of 80 LEO satellites).

Today, only a minority of cable operators continue to enjoy any *de facto* monopoly in the provision of residential high-speed Internet access. The majority, in contrast, are now subject to intermodal competition, especially from DSL.[140] In just two years, these intermodal competitors have captured 32% of the still-nascent market.[141] On the other hand, cable continues to enjoy a number of advantages over its intermodal competitors, including both the cost savings that cable operators derive from the incumbency of their existing "last mile" facilities, and the technological capacity of HFC cable to offer substantially higher transmission speeds than competing facilities.

### V. Two Controversies Over Access.

Because local cable television systems and international telecommunications satellites each were originally conceived of as "natural monopolies," these two seemingly dissimilar communications media inspired parallel regulatory responses. Both types of facilities initially enjoyed legal protection against intramodal competition, and were protected by circumstances against effective intermodal competition. Both were initially subject to substantial regulation of rates and service offerings. Neither, however, were subject to access or unbundling requirements. Eventually, in both cases, longstanding bans on intramodal competition were lifted, while concurrent technological advances enhanced the prospects for effective intermodal competition. These transitions to conditions of competition were accompanied in both cases by relaxation of previously applicable rate and service regulation. Consistent with this transition, when cable modem service was introduced by cable operators in the late 1990s, the new service was never subject to rate and service regulation or to mandatory unbundling or competitive access requirements. To varying extents, the classical telecommunications paradigm of the "regulated monopoly" gave way to the modern ideal of "market competition" in the local cable and international satellite sectors alike.

---

[140] *See In re Inquiry Concerning High-Speed Access to the Internet Over Cable and Other Facilities, Declaratory Ruling & NPRM*, FCC 02-77, GEN Docket No. 00-185, 2002 WL 407567, ¶ 9 n.25 (rel. Mar. 15, 2002) (noting that 57.5% of U.S. Zip Codes are served by more than one facilities-based provider of residential high-speed Internet access; 20.3% are served by only one; and 22.2% are served by none).

[141] *In re Inquiry Concerning High-Speed Access to the Internet Over Cable and Other Facilities, Declaratory Ruling & NPRM*, FCC 02-77, GEN Docket No. 00-185, 2002 WL 407567, ¶ 9 (rel. Mar. 15, 2002); *accord* U.S. Department of Commerce, National Telecommunications and Information Administration, *A Nation Online: How Americans Are Expanding Their Use of the Internet* 35 (Feb. 2002), available online at <http://www.ntia.doc.gov/ntiahome/dn/anationonline2.pdf>. *But see* James B. Speta, *A Common Carrier Approach to Internet Interconnection*, 54 Fed. Comm. L.J. 225, 234 (2002) (asserting that cable modem service has enjoyed a "continued *and expanding* lead . . . over its main high-speed competitor, DSL service, at least in residential markets.") (emphasis added) (citing law review articles).

A mutual transition from monopoly to competition, however, was not the only attribute shared by the local cable television and international satellite telecommunications industries. Rather, the two industries had something else in common: even as they were generally being deregulated, both industries experienced parallel controversies over customers' or competitors' demands for new regulation that would require mandatory unbundling of services and facilities. The first of these controversies concerned whether unaffiliated U.S. international common carriers should enjoy a legal right to bypass INTELSAT's U.S. retail affiliate (COMSAT) and purchase raw satellite transmission capacity directly from INTELSAT on the same terms and conditions that such capacity could be purchased by COMSAT. The second controversy concerned whether unaffiliated ISPs should enjoy a legal right to bypass the retail "cable ISP" affiliates of facilities-based cable system operators and purchase raw cable transmission capacity directly from the cable system operator on the same terms and conditions that such capacity could be purchased by the affiliated cable ISP. The history of each of these controversies is discussed in this Part.

### A. The "INTELSAT Direct Access" Debate.

Until 1985, non-INTELSAT satellites were not permitted to provide international communications services to or from the United States.[142] At the same time, INTELSAT's U.S. affiliate COMSAT[143] enjoyed an exclusive franchise over the

---

[142] *Cf. Establishment of Satellite Systems Providing International Communications*, 101 F.C.C. 2d 1046 (1985) ("*Separate Systems Order*") (authorizing satellite systems separate from INTELSAT to provide international telecommunications service to and from the United States), *modified on recon.*, 61 Rad. Reg. (P&F) 649 (1986), *further recon. denied*, 1 FCC Rcd 439 (1986).

[143] The national governments that entered into the INTELSAT Agreement are referred to as "Parties" to the INTELSAT treaty organization. *See Agreement Relating to the International Telecommunications Satellite Organization "INTELSAT,"* Art. I, done Aug. 20, 1971, 23 U.S.T. 3813. The Agreement, however, does not require its member state governments to assume an active role in financing or operating the satellite system, but instead requires each Party to "designate a telecommunications entity, public or private," to assume responsibility for financing its country's share of the satellite system and for performing certain commercial operations. *Id.* Art. II(b). These designated telecommunications entities, which, until privatization, collectively owned and operated the satellite facilities of the INTELSAT system, were known as "Signatories" to the INTELSAT Operating Agreement, a separate commercial agreement signed by the "Signatories" (and not by the State Parties) that established procedures for governing INTELSAT's commercial operations. *See Operating Agreement Relating to the International Telecommunications Satellite Organization, "INTELSAT,"* done Aug. 20, 1971, 23 U.S.T. 4091. In the United States, COMSAT Corp., a publicly traded District

provision of INTELSAT satellite space segment capacity to carriers and users located in the United States.[144] For these reasons, since at least the 1970s, several U.S. international common carriers who were then COMSAT's principal customers sought to require INTELSAT to unbundle its "raw" satellite "space segment" transmission capacity, and to sell that space segment capacity to unaffiliated entities on the same terms and conditions under which it sold the capacity to COMSAT.[145] This proposal for mandatory

---

of Columbia corporation, was designated as the "U.S. Signatory" to INTELSAT. *Direct Access To The INTELSAT System, Report & Order*, 14 FCC Rcd 15703, ¶ 5 (1999). Each Signatory's investment share in INTELSAT was required to remain in proportion to its share of the total use of INTELSAT satellite transmission capacity. INTELSAT Agreement Art. V(b). Accordingly, in 1999, when the FCC implemented direct access to INTELSAT, COMSAT's affiliation with INTELSAT was manifested in its 20.4% investment share in INTELSAT, the largest of any Signatory. COMSAT Corp. 1999 SEC Form 10-K, at 6 (filed Mar. 30, 2000), *online at* <http://www.sec.gov/Archives/edgar/data/22698/0000928385-00-001038-index.html>.

[144] Under the FCC's "*Authorized User I*" policy (in effect from 1966 to 1984), COMSAT was deemed a "carrier's carrier," authorized to furnish INTELSAT capacity *only* to FCC-designated common carriers, and forbidden from furnishing any INTELSAT services or capacity directly to end users except in certain special circumstances. *See Authorized User I*, 4 F.C.C. 2d 421 (1966). During the two decades that this policy was in effect, however, the FCC recognized an ever-increasing number of such "special circumstances" in which COMSAT could serve non-carrier end users directly. *See*, *e.g.*, *Spanish Int'l Network*, 70 F.C.C.2d 2127 (1978) (authorizing COMSAT to provide international television program transmission service directly to non-carrier users). Finally, in 1984, the FCC formally rescinded the *Authorized User I* policy, and authorized COMSAT to bypass the international common carriers and begin providing end-to-end international communications services directly to certain "authorized" large non-carrier end users (such as television broadcast networks). *See Modification of Authorized User Policy*, 90 F.C.C. 2d 1394 (1982) ("*Authorized User II*"), *vacated & remanded*, *ITT World Communications, Inc. v. FCC*, 725 F.2d 732 (D.C. Cir. 1984), *and reinstated on remand*, 98 F.C.C. 2d 158 (1984).

Confusingly, the FCC initially characterized its new *Authorized User II* policy as "authoriz[ing] non-carrier users to gain *direct access* to Comsat's INTELSAT basic transmission services." *Id.* ¶ 69. (emphasis added). This policy, however, should not be confused with "direct access to INTELSAT," which—unlike the *Authorized User II* policy--authorizes carriers *and* non-carrier users to bypass COMSAT, and to obtain INTELSAT space segment capacity directly from INTELSAT. *See* note **[n+2]**, *infra*. During the period from 1984 to 1999 during which the *Authorized User II* policy was in effect but "direct access" was not, U.S. common carriers could obtain the INTELSAT space segment capacity only from COMSAT, as part of a bundled service offering.

[145] *See, e.g., In re Application of Communications Satellite Corp. For Authority To Provide Satellite Television Services Directly to Users at United States Earth Stations*, 76

(continued . . .)

unbundling of transmission capacity from telecommunications service was denoted "direct access to INTELSAT."[146]

Although earlier proposals for direct access had stalled, the issue was revived in 1982 when the FCC authorized COMSAT, for the first time, to bypass U.S. international common carriers (*e.g.*, AT&T, ITT, Western Union), and to provide end-to-end international communications services directly to "authorized" large non-carrier end

---

F.C.C. 2d 5, ¶ 6 (1980) (acknowledging that several international common carriers had proposed to "be allowed direct access to the INTELSAT operations center for the purpose of placing orders for television service, or that carriers be allowed to acquire satellite facilities on an equal basis with Comsat either on an indefeasible right of user basis or by directing that such facilities be provided on a cost-pass-through basis."). As early as 1980, the Commission noted that "[t]he issue of direct carrier access to Intelsat is not new. . . . The [carriers] frequently raise the issue of direct access." *Id.* ¶¶ 22-23.

[146] *See also Direct Access To The INTELSAT System, Report & Order*, 14 FCC Rcd 15703, ¶ 1 (1999) ("'Direct access' refers to the means by which users of the INTELSAT satellite system may obtain space segment capacity directly from INTELSAT rather than having to go through an INTELSAT Signatory"). When discussing "direct access to INTELSAT," commentators sometimes use the words "unbundling" and "direct access" interchangeably. *See, e.g.*, Robert M. Frieden, *International Telecommunications and the Federal Communications Commission*, 21 Colum. J. Transnat'l L. 423, 459 (1983) (arguing that "[t]he Commission must ensure that carriers other than Comsat can secure direct and cost-based access to INTELSAT space segment on the same terms and conditions that Comsat secures. Toward that end, Comsat should be directed to unbundle the INTELSAT space segment cost from associated earth station services and facilities expenses.") (footnote omitted).

In fact, INTELSAT offers four different types, or "levels," of direct access to unaffiliated carriers. *Direct Access To The INTELSAT System,* 14 FCC Rcd 15703, ¶ 8 (1999). This Paper, like the FCC's direct access orders and the ORBIT Act, uses the phrase "direct access" as a term of art to denote the unbundling of INTELSAT space segment capacity that INTELSAT formally classifies as "Level III Direct Access." *See id.* ¶¶ 2, 8 ("Level 3 direct access permits a customer to enter into a contractual agreement with INTELSAT for the purpose of ordering, receiving, and paying for INTELSAT space segment capacity at the same rates that INTELSAT charges its Signatories."). For a description of the other types of INTELSAT direct access not at issue here, *see generally id.* ¶ 8 (summarizing *Accessing INTELSAT . . . Directly*, *reprinted in* Record of Hearing Before the House Subcomm. on Telecommunications, Trade, and Consumer Protection on H.R. 1872, 105[th] Cong., at 135-141, and INTELSAT AP-21-E *Report by the Board of Governors on INTELSAT Access Arrangements*, March 18, 1997). *See also* Charles H. Kennedy & M. Veronica Pastor, *An Introduction to International Telecommunications Law* 74-75 (1996) (same).

users (such as television broadcast networks).[147]  This policy led several major U.S.-international common carriers to renew their call for "direct access" to INTELSAT. These carriers argued that if COMSAT is permitted to bypass the U.S. carriers in order to sell INTELSAT capacity directly to end users, then the U.S. carriers should correspondingly be permitted to bypass COMSAT and purchase raw satellite capacity directly from INTELSAT on the same terms and conditions as COMSAT.[148]  To address the carriers' concerns, the FCC initiated a new Notice of Inquiry on "direct access."[149]  In this Notice, the FCC noted that the carriers (and the U.S. Justice Department and National Telecommunications and Information Administration) had argued that unbundled "direct access" to INTELSAT would serve the public interest by:

- enhancing price competition among international telecommunications carriers by enabling those carriers to compete effectively against COMSAT in the market for provision of INTELSAT space segment capacity and transmission services to end users;[150]

- lowering the price paid by international carriers to acquire INTELSAT space segment capacity (causing such savings to be passed through to consumers);[151]

- guarding against COMSAT's using its monopoly INTELSAT role to engage in unfair practices (such as predatory underpricing of services offered directly to the public) in competitive markets;[152]

- permitting (and inspiring) additional carrier investment in INTELSAT satellite facilities, and thereby enhancing deployment of such facilities;[153] and

---

[147]   *Authorized User II*, 90 F.C.C. 2d 1394 (1982).

[148]   *See id.* at ¶¶ 100-105.  The international common carriers use the satellite capacity to carry international voice and data calls.

[149]   *See Regulatory Policies Concerning Direct Access to INTELSAT Space Segment for the U.S. International Service Carriers, Notice of Inquiry* ("*1982 Direct Access NOI*"), 90 F.C.C. 2d 1446 (1982), *dismissed*, 97 F.C.C. 2d 296 (1984), *dismissal aff'd*, *Western Union Int'l, Inc. v FCC*, 804 F.2d 1280 (D.C. Cir. 1986).

[150]   *Id.* ¶¶ 1, 9; *see also Regulatory Policies Concerning Direct Access to INTELSAT Space Segment for the U.S. International Service Carriers*, *Report and Order*, 97 FCC 2d 296, ¶¶ 11, 56 (1984) ("*1984 Direct Access Order*"), *aff'd Western Union Int'l, Inc. v FCC*, 804 F.2d 1280 (D.C. Cir. 1986).

[151]   *1982 Direct Access NOI*, 90 F.C.C. 2d 1446, ¶ 2 (1982).

[152]   *Id.* ¶¶ 7, 12-13; *see also 1984 Direct Access Order*, 97 FCC 2d 296, ¶¶ 12, 56 (1984).

- fostering innovation and efficiency in the provision of international communications services.[154]

As discussed in Part V, *infra*, these arguments advanced in the early 1980s in favor of carrier "direct access" to INTELSAT are substantially similar to the arguments currently being advanced in favor of ISP "open access" to HFC cable capacity used for high-speed data transmission.

In 1984, the FCC terminated the 1982 "direct access" proceeding after concluding that "proponents of direct access have failed to establish that it will serve the public interest."[155] At that time, the Commission was "unpersuaded that, whatever benefits are to be derived, they would be so substantial as to outweigh the adverse consequences which are likely to attend the adoption and implementation of direct access."[156] Specifically, the Commission concluded, *inter alia*, that:

- Neither INTELSAT's underlying costs of operating and maintaining the global satellite system nor the corresponding wholesale IUC rates that it charged would likely be affected by the implementation of "direct access" in the United States.[157]

---

[153] *1982 Direct Access NOI*, 90 F.C.C. 2d 1446, ¶ 8 (1982); *see also 1984 Direct Access Order*, 97 FCC 2d 296, ¶¶ 14-15 (1984).

[154] *1982 Direct Access NOI*, 90 F.C.C. 2d 1446, ¶ 9.

[155] *1984 Direct Access Order*, 97 FCC 2d 296, ¶ 3 (1984). The Commission issued the *1984 Direct Access Order* shortly after the D.C. Circuit court ordered it to conclude its still-open 1982 "direct access" proceeding. *See ITT World Communications, Inc. v. FCC*, 725 F.2d 732, 754-55 (D.C. Cir.1984) (affirming the FCC's *Authorized User II* Order, but directing the FCC to give due consideration to the carriers' requests for "direct access" to INTELSAT).

[156] *1984 Direct Access Order*, 97 FCC 2d 296, ¶ 3 (1984); *see also id.* ¶ 32 (concluding that carrier direct access to INTELSAT "would not produce significant economic savings to carriers or users").

[157] *1984 Direct Access Order*, 97 FCC 2d 296, ¶¶ 40, 44 (1984). INTELSAT's "wholesale" charges for use of raw satellite transmission capacity are termed "INTELSAT Utilization Charges" ("IUCs"). "The IUC serves as a measure, expressed on a per-circuit basis, of the costs INTELSAT incurs in constructing and operating the global satellite system." *Id.* ¶ 34. "The level of the IUC is set so as to recover amortization of capital (depreciation expense), the operating expenses of the INTELSAT system and compensation to Signatories for the use of their capital." *Id.* The IUC does not reflect the local "internal costs" of connecting a carrier or user to an INTELSAT satellite circuit. *Id.*

- COMSAT's costs of administering INTELSAT programs and establishing INTELSAT circuit connectivities in the United States—which were not recovered in INTELSAT's wholesale IUC rates—would not be avoided by implementation of "direct access," and would need to be recovered elsewhere.[158]

- Because even COMSAT's new "end-to-end" service would furnish only the communications link from a U.S. earth station to an INTELSAT satellite—but would *not* supply the terrestrial end-links between the earth station and the end user's premises—COMSAT did not enjoy any *intramodal* competitive advantage over U.S. carriers.[159]

- Because transoceanic submarine cables were capable of providing many of the same services that INTELSAT could provide, and at lower prices, COMSAT was always subject to *intermodal* competition.[160]

---

[158] *Id.* ¶¶ 44-47. Such costs include "COMSAT's return on its satellite investment; the income tax Comsat is obligated to pay on that return; and Comsat's internal operations and maintenance, research and development, and corporate headquarters expenses." *Id.* ¶ 44. With respect to such costs, the Commission concluded in 1984 that "were we to adopt direct access, we would merely be changing the form in which these expenses would be recovered and, in the process, adding an unnecessary layer of regulation with its attendant costs." *Id.* ¶ 50.

[159] *Id.* ¶ 58. Specifically, the FCC concluded that although the U.S. common carriers might be competitively disadvantaged in having to obtain the earth-station-to-satellite link from COMSAT at "retail" price, the carriers possessed a compensating competitive advantage in consequence of being able to furnish terrestrial ground-links to end users, which COMSAT could not supply. *Id.* Stated differently, the Commission concluded that the carriers' ability to offer retail "one-stop-shopping" to end users for a panoply of international satellite communications services provided effective competition to COMSAT's ability to obtain a single network element of such service (the satellite link) at the "wholesale" IUC rate. *See id.* at ¶¶ 58-62 (providing examples in which end users preferred to take seamless end-to-end service from a non-COMSAT carrier).

[160] *Id.* ¶¶ 63-64. In 1984, the communications behemoth AT&T owned virtually all transoceanic submarine cables to or from the United States, while also utilizing 90% of the INTELSAT space segment capacity then leased to U.S. carriers. *Id.* ¶ 64. In consequence of this fact, the FCC noted that "the effect on intermodal competition of . . . direct access . . . would depend to a significant degree on the behavior of AT&T." *Id.* While the Commission could not predict exactly what AT&T (which opposed "direct access" to INTELSAT) would actually do, it did note that under a "direct access" regime, "AT&T could in effect control investment decisions relating to INTELSAT space segment, by and large removing that function from Comsat. Such a development could be detrimental to the promotion of intermodal competition, insofar as AT&T could bias investment and circuit utilization decisions in favor of one medium over the other." *Id.*

- Even assuming that direct access would produce savings for carriers, and that these savings would be passed-through dollar-for-dollar by carriers to end-users, such savings would be imperceptible to end-users, because they would not constitute more than a few percentage points of the total end-user charge.[161]

Notably, because it found that direct access to INTELSAT would not serve the public interest, the *1984 Direct Access Order* did not resolve whether the FCC possessed legal authority to implement "direct access" to INTELSAT.[162]

The issue of "direct access to INTELSAT" next resurfaced in the mid-1990s, when COMSAT petitioned the FCC for reclassification as a non-dominant carrier.[163] By that time, as discussed in Part III.B, *supra*, "several U.S. separate satellite systems provide[d] service to many foreign nations."[164] Moreover, the early 1990s saw an unprecedented deployment of transoceanic submarine cables capable of providing many of the same services that INTELSAT provides, at lower cost.[165] In response to COMSAT's petition, several of COMSAT's customers and competitors commented that COMSAT should be reclassified as non-dominant only in connection with the implementation of direct access to INTELSAT in the United States.[166]

---

[161] *Id.* ¶¶ 66-67. In 1984, COMSAT obtained raw space segment capacity from INTELSAT at an IUC rate of $390 per circuit per month, and incurred a total cost (including expenses) of $800 per circuit per month to obtain such capacity. *Id.* ¶¶ 36, 38. COMSAT then leased these circuits to U.S. carriers and users at a rate of $1125 per circuit per month, a rate designed to recover both its cost of obtaining the INTELSAT space segment capacity, and its additional cost of operating its proprietary earth station facilities, which receive the INTELSAT signals. *Id.* After taking possession of the signals at the COMSAT earth station, the domestic carriers used their own terrestrial facilities to transmit the signal to the premises of end-users located in the United States. *Id.* ¶ 66. "The typical carrier charge [to an end-user] for a voice-grade (telephone) end-to-end service is approximately $4,000 per channel per month." *Id.* Based on these figures, the FCC concluded that end-users would receive only a minimal benefit even if the carriers could obtain raw INTELSAT space segment capacity at a price closer to the IUC. *Id.*

[162] *See id.* at ¶¶ 27-29 (summarizing arguments advanced by COMSAT that "direct access" was prohibited by the Communications Satellite Act of 1962, but declining to address such arguments). In affirming the *1984 Direct Access Order*, the D.C. Circuit endorsed the FCC's policy conclusions, and likewise declined to address whether the Commission would have had legal authority to require "direct access" had it desired to do so. *See Western Union Int'l, Inc. v. FCC*, 804 F.2d 1280, 1285-87 (D.C. Cir. 1986).

[163] *See In re COMSAT Corp. Petition for Reclassification as a Non-Dominant Carrier*, 13 FCC Rcd 14083 (1998) ("*COMSAT Non-Dominant Order*"), *supplemented*, *Policies & Rules for Alternative Incentive Based Regulation of COMSAT Corp.*, 14 FCC Rcd 3065 (1999).

[164] *Id.* ¶ 11.

Without implementing direct access, the *COMSAT Non-Dominant Order* granted COMSAT's request for reclassification as a non-dominant carrier on every international route over which INTELSAT was subject to facilities-based competition.[167] At the same time, however, the *Order* called upon COMSAT to voluntarily allow direct access to INTELSAT on the few remaining non-competitive "thin routes" not served by any satellite or cable other than INTELSAT.[168] According to the *Order*, permitting unaffiliated carriers and users to obtain "direct access" to INTELSAT on the noncompetitive U.S. international routes would serve the public interest by:

- reducing COMSAT's control in the U.S. over the supply of INTELSAT satellite capacity serving non-competitive markets, and thereby reducing COMSAT's market power in these markets;

---

[165] *Id.*

[166] *Id.* ¶ 17 & n.49. The repeated efforts of COMSAT's customers and competitors to raise the issue of direct access to INTELSAT in the context of FCC regulatory proceedings initiated by COMSAT parallels the similar efforts of cable ISP customers and competitors to raise the "cable open access" issue in the context of FCC (and local government) regulatory proceedings (such as the proceedings seeking regulatory approval of the AT&T/TCI Cable merger and the AOL/Time Warner merger) initiated by the cable companies. In both situations, access proponents have argued that the regulatory result sought by the respective applicants should be granted only if conditioned on the pertinent access requirement. *See* note __, *infra* (discussing cable open access debates that arose in the FCC's proceedings on the AT&T/TCI, AT&T/MediaOne, and AOL/Time Warner mergers).

[167] Specifically, the *COMSAT Non-Dominant Order* reclassified COMSAT as a non-dominant telecommunications communications carrier in the switched-voice, private-line, and occasional-use video markets on every major route to or from the United States. COMSAT was *not*, however, reclassified as non-dominant of the few remaining "thin routes" not served by any other satellite system or submarine cable. On those routes (which accounted for about 7% of COMSAT's INTELSAT revenues in 1999), COMSAT remains subject to dominant carrier tariffing requirements and price-cap rate regulation. *See Policies & Rules for Alternative Incentive Based Regulation of COMSAT Corp.*, 14 FCC Rcd 3065 (1999) ("*Thin Route Order*") (setting forth the FCC's regulatory policies concerning COMSAT's remaining "thin routes").

[168] *See COMSAT Non-Dominant Order*, 13 FCC Rcd 14083, ¶¶ 155-56 (1998).

- affording U.S. international carriers and users a choice between using Comsat or accessing INTELSAT directly to serve the non-competitive markets;

- spurring competition among telecommunications service providers, and thereby promoting competitive market conditions in markets that were then non-competitive; and

- creating the potential for price competition, service quality improvements and innovation.[169]

The *COMSAT Non-Dominant Order* did not discuss any public interest benefits that might accrue if direct access to INTELSAT were implemented on the *competitive* routes that accounted for 93% of COMSAT's INTELSAT revenues, Nonetheless, a new rulemaking proceeding was soon initiated to reconsider implementing direct access to INTELSAT, without limitation to noncompetitive routes.[170] One year later, after decades of resisting similar proposals, the Commission adopted its latest direct access proposal.[171] Finally, on March 17, 2000, Congress enacted the comprehensive ORBIT Act, which codified the *1999 Direct Access Order* by expressly providing for direct access to INTELSAT in the United States.[172] Today, the FCC continues to play an active role in the

---

[169] *Id.* ¶ 155.

[170] *Direct Access To The INTELSAT System, Notice of Proposed Rulemaking*, 13 FCC Rcd 22013 (1998). The FCC announced its intention to initiate this new rulemaking proceeding in the *COMSAT Non-Dominant Order*, 13 FCC Rcd 14083, ¶¶ 156-57 (1998). In so doing, it noted that in 1984 (when the FCC had earlier declined to adopt direct access to INTELSAT in the United States), direct access to INTELSAT had not been implemented anywhere on earth, and INTELSAT then had no procedures for implementing it. In 1992, however, "INTELSAT introduced new procedures for gaining direct access to INTELSAT satellites by non-Signatory carriers and users." *1999 Direct Access Order*, 14 FCC Rcd 15703, ¶ 8 (1999). By 1998, when COMSAT was reclassified as non-dominant, direct access to INTELSAT had been adopted in 76 foreign countries, including the United Kingdom. *COMSAT Non-Dominant Order*, 13 FCC Rcd 14083, ¶ 157 (1998).

[171] *Direct Access To The INTELSAT System, Report & Order*, 14 FCC Rcd 15703 (1999) ("*1999 Direct Access Order*").

[172] *See* Open-Market Reorganization for the Betterment of International Telecommunications (ORBIT) Act § 641(a), Pub. L. No. 106-180 § 641(a), 114 Stat. 48, 55 (2000), *codified at* 47 U.S.C. §§ 765(a) (Supp. 2001) (providing that "users or providers of telecommunications services shall be permitted to obtain direct access to INTELSAT telecommunications services and space segment capacity through purchases of such capacity or services from INTELSAT."). Section 641(a) of ORBIT did nothing more than codify the FCC's *1999 Direct Access Order* (which had been issued six months earlier). However, both Congress and the FCC considered enactment of the

(continued . . .)

transition to "direct access" to INTELSAT. Indeed, the Commission is currently considering a proposal to abrogate some of INTELSAT's existing contractual capacity obligations in order to free up more capacity for "direct access" users.[173]

### B. The Cable "Open Access" Debate

Before residential cable modem service was deployed in 1998, most residential users had no means of obtaining Internet access from their homes other than "via traditional 'dial-up' telephone services provided by local exchange carriers (LECs) over copper telephone lines."[174] In those days, most LECs enjoyed monopolies over the provision of dial-tone service to their customers' homes. Because, however, LECs were required by law to unbundle dial-tone (transportation) service from ISP service, the LECs' monopoly over dial-tone service did not extend into the ISP market.[175] Rather, these unbundling requirements provided residential consumers with a choice of whether to take ISP service from an ISP affiliated with their LEC, or through a competing unaffiliated ISP.[176]

The legal obligation of LECs to unbundle their transportation services from their ISP services derived from several sources. First, as regulated "public utility corporations," LECs had never been permitted to deny telephone service to anyone willing to pay, including a potential competitor.[177] Thus, LECs could not prevent unaffiliated ISPs from obtaining

---

provision to be necessary because COMSAT's petition for review of the *1999 Direct Access Order*, which was then pending before the D.C. Circuit, raised substantial questions about the Commission's legal authority under the Communications Satellite Act of 1962 to implement direct access unilaterally. To resolve these questions, the "direct access" provision was attached to the comprehensive ORBIT Act, whose main provisions mandated a mechanism for privatization of INTELSAT and Inmarsat.

[173] *See In re Availability of INTELSAT Space Segment Capacity To Users and Service Providers Seeking To Access INTELSAT Directly*, 15 FCC Rcd. 19160, ¶¶ 2, 5 (2000) ("*INTELSAT Capacity Order*").

[174] *Notice of Inquiry Concerning High-Speed Access to Internet*, 15 FCC Rcd. 19287, ¶ 6 (2000).

[175] *See Notice of Inquiry Concerning High-Speed Access to Internet*, 15 FCC Rcd. 19287, ¶ 6 (2000). ISP service differs from transportation service; "[i]n general, ISPs receive communications from their customers' computers and route the communications to other computers connected either to their networks or other networks." *Id.* ¶ 9.

[176] *See id.*

[177] *See, e.g.*, *North Carolina Pub. Serv. Co. v. Southern Power Co.*, 282 F. 837, 844 (4th Cir. 1922), *cert. dismissed*, 263 U.S. 508 (1924) ("when a [public utility] corporation
<space style="margin-left: 4em">(continued . . .)</space>

<space style="margin-left: 2em">46</space>

telephone numbers. Similarly, as regulated "public utility corporations," LECs could not require a consumer to purchase an unregulated service in order to obtain the regulated service.[178] This longstanding regulatory doctrine foreclosed the LECs from "bundling" an affiliated ISP service with their standard local calling service, and requiring consumers to take both or neither.

In addition, certain provisions of the Telecommunications Act of 1996 that were designed to jumpstart competition in local calling markets now require most LECs "to 'unbundle' and sell to their competitors whatever *new* capabilities they add to their networks at rates 'based on the cost[s] of providing" them."[179] Accordingly, "[b]y virtue of the local competition provisions of the Telecommunications Act of 1996, [incumbent LECs now] must grant unbundled access to competing carriers seeking to provide DSL service."[180] In addition, "[a] cluster of FCC rules that survived the passage of the 1996 Act similarly entitles unaffiliated [ISPs] to request interconnection and unbundled sale of network elements from the largest ILECs for the purpose of providing DSL service."[181]

---

has definitely undertaken and entered upon a particular service authorized by a charter . . ., the obligation to perform the service is complete, its rates and terms are subject to regulation by public authority, and it must serve all alike. In such public service it cannot pick and choose its customers."). The traditional duties of telecommunications common carriers to interconnect and to provide service to their competitors at just and reasonable rates is reflected in the Communications Act of 1934, see 47 U.S.C. §§ 201, 203-205.

[178] This result derived from longstanding regulatory doctrines which barred public utility corporations from bundling regulated monopoly services with services also available from competitors. *See, e.g.*, *Hicks v. City of Monroe Utilities Comm'n*, 112 So.2d 635, 647 (La. 1959) (a "public utility corporation[] cannot refuse to render the service which it is authorized by its charter to furnish because [of a customer's refusal to purchase a different service] . . . Each kind of service must be furnished on its own merits and no discrimination is permitted against a customer for one service because he does not desire another service"); *Seaton Mountain Elec. Light, Heat & Power Co. v. Idaho Springs Inv. Co.*, 111 P. 834, 835-46 (Colo. 1910) (public utility corporation could not condition its provision of steam service on customer's additional purchase of electric service).

[179] Peter W. Huber, Michael W. Kellogg, & John Thorne, *Federal Telecommunications Law* § 11.9.2, at 1070 (2d ed. 1999) (emphasis added) (quoting 47 U.S.C. § 251(c)(3) and 47 U.S.C. § 252(d)(1)(A)(i)).

[180] Jim Chen, *The Authority To Regulate Broadband Internet Access Over Cable*, 16 Berkeley Tech. L.J. 677, 680 & n.18 (2001) (citing FCC Orders implementing 47 U.S.C. § 251(c)(3)).

[181] *Id.* at 680-81 & n.19 (citing FCC Orders and a treatise). *See generally* Peter W. Huber, Michael W. Kellogg, & John Thorne, *Federal Telecommunications Law* §§ 12.5.2-.3, at 1097-1103 (2d ed. 1999) (explaining the significance and ongoing vitality of

Thus, both the legacy of public utility regulation and the local competition provisions of the Telecommunications Act of 1996 have brought "open access" (*i.e.* unbundling) regulatory requirements to all forms of Internet access delivered to residential users by telephone wires, from "narrowband" to DSL.[182] At the same time, the separate legacy of cable television regulation (including the modern trend towards cable television *de*regulation) has enabled cable modem service providers to avoid being subject to similar requirements.[183] Unlike incumbent LECs offering dial-up or high-speed DSL service, "cable operators currently are not legally prohibited from having an exclusive relationship with one particular ISP," including an affiliated or wholly-owned ISP.[184] Primarily because of these legacy-based regulatory distinctions, cable modem service providers are now immune from unbundling and access requirements that apply to DSL service providers, even though the two services are fungible to the consumer.[185]

---

the "Expanded Interconnection" and "Open Network Architecture" rules); James B. Speta, *Handicapping the Race for the Last Mile?: A Critique of Open Access Rules for Broadband Platforms*, 17 Yale J. on Reg. 39, 67-69 (2000) (detailing the legal obligation of ILECs to deal with competing DSL carriers).

[182] *See In re Appropriate Framework for Broadband Access to the Internet Over Wireline Facilities*, *Universal Service Obligations of Broadband Providers*, *Notice of Proposed Rulemaking*, FCC 02-42, CC Docket No. 02-33, 2002 WL 252714, ¶¶ 6-7 ("*Wireline Broadband NPRM*") (rel. Feb. 15, 2002), *reprinted in* 67 Fed. Reg. 9232 (Feb. 28, 2002).

[183] *See generally* Rosemary Harold, **Cable-Based Internet Access: Exorcising the Ghosts of "Legacy" Regulation,** 28 N. Ky. L. Rev 721 (2001) (discussing how telephony's legacy of common carrier interconnection and access requirements, and cable television's historical lack thereof, have contributed to the divergence between the respective access requirements now applied to DSL and cable ISPs); see also *See In re Appropriate Framework for Broadband Access to the Internet Over Wireline Facilities*, *Universal Service Obligations of Broadband Providers*, *Notice of Proposed Rulemaking*, FCC 02-42, CC Docket No. 02-33, 2002 WL 252714, ¶ 6 ("*Wireline Broadband NPRM*") (rel. Feb. 15, 2002) ("Legacy regulations were based on technical and market assumptions concerning [cable and telephone] networks and the services they delivered."), *reprinted in* 67 Fed. Reg. 9232 (Feb. 28, 2002).

[184] *Notice of Inquiry Concerning High-Speed Access to Internet*, 15 FCC Rcd. 19287, ¶ 29 (2000). Like cable television service, neither fixed wireless telecommunications services nor DBS television service have a legacy of common carrier regulation. Accordingly, like cable but unlike DSL, residential ISP service using terrestrial or satellite wireless high-speed transmission technologies are not subject to "open access' or "unbundling" requirements.

[185] *See In re Appropriate Framework for Broadband Access to the Internet Over Wireline Facilities*, *Universal Service Obligations of Broadband Providers*, *Notice of*

Because of the apparent arbitrariness of the distinction, many unaffiliated ISPs and some public interest groups have advocated that cable systems offering high-speed data transmission services should be subject to "an 'open access' requirement similar to the common carrier regime under Title II of the Communications Act," which would require cable operators who offer high-speed data transmission services "to grant unaffiliated ISPs non-discriminatory access to their cable plant."[186] "These groups argued that broadband service over cable lines is essentially common carriage, and moving bits between an ISP and a consumer is essentially a transmission service."[187] Cable open access proponents have also asserted that an "open access" requirement would speed deployment of high-speed Internet service to the residential market and produce other competitive benefits;[188] protect the end-to-end architecture of the Internet;[189] or that mandatory unbundling cable ISP services from cable transportation services is needed to forestall future regulation of both of those services.[190]

---

*Proposed Rulemaking*, FCC 02-42, CC Docket No. 02-33, 2002 WL 252714, ¶ 7 ("*Wireline Broadband NPRM*") (rel. Feb. 15, 2002) (recognizing that legacy regulation has led to divergent regulatory requirements now being applied to cable modem and DSL service despite the fact that these two platforms provide substantially the same service to the consumer), *reprinted in* 67 Fed. Reg. 9232 (Feb. 28, 2002).

[186]   Deborah A. Lathen, *Broadband Today, FCC Staff Report on Industry Monitoring Sessions* 36 (1999), *available online at* <http://www.fcc.gov/Bureaus/Cable/Reports/broadbandtoday.pdf>.

[187]   *Id.*; *accord* James B. Speta, *Handicapping the Race For the Last Mile? A Critique of Open Access Rules For Broadband Platforms*, 17 Yale J. on Reg. 39, 78 (2000) (The "arguments for open access . . . have a strong pedigree in the history of telecommunications regulation, which has long been concerned that dominant telephone companies will use their control over transport facilities to act anti-competitively against computer companies or other information service providers.").

[188]   See Part IV.A-IV.B, *infra* (discussing debates over whether access requirements spur competition and deployment).

[189]   *See* Mark A. Lemley & Lawrence Lessig, *The End of End-To-End: Preserving the Architecture of the Internet in the Broadband Era*, 48 UCLA L. Rev. 925, 971 (2001).

[190]   *See* Mark A. Lemley & Lawrence Lessig, *The End of End-To-End: Preserving the Architecture of the Internet in the Broadband Era*, 48 UCLA L. Rev. 925, 971 (2001) (asserting that "if we let natural monopoly services be bundled together with potentially competitive services, we will end up having to regulate not only the monopoly services but the competitive ones as well. . . . [Ironically], to allow cable companies to tie their natural monopoly service to a competitive one . . . will lead in the end to more regulation, and what is worse, to unnecessary regulation.").

Similarly, most LECs also argue for regulatory parity between cable and DSL.[191] Unlike the unaffiliated ISPs, however, the LECs suggest that regulatory parity should be achieved *not* by imposing "open access" on cable ISPs, but instead by eliminating the "open access" and "unbundling" requirements that currently apply to DSL.[192]

In response to the arguments advanced by "cable open access" proponents, some local governments attempted to impose "open access" requirements upon cable television franchisees in connection with transfers or renewals of their franchises.[193] To date, however, the two U.S. Circuit Courts of Appeals that have reviewed such local ordinances both held the entire field to be preempted by federal law.[194] At the same time, however, these two circuit courts both opined that the federal law at issue vested the FCC with authority either to require "cable open access" or to prohibit it, and that the FCC had not exercised this authority one way or the other.[195] In so holding, these courts returned the "cable open access" debate squarely to the doorstep of the FCC.[196]

---

[191] *See, e.g.*, Comments of SBC Corp. and BellSouth Corp *filed in* FCC GEN Docket No. 00-185, at i (filed Dec. 1, 2000) (asserting that "like services must be treated alike, regardless of the name, corporate history, or traditional lines-of-business of the service provider. Broadband Internet access is the same service, whether it is provided over coax, over copper, or through the air.").

[192] *See, e.g., id.* at iii ("Incumbent LECs should stand on equal footing with other [high-speed transmission] service providers, . . . [who] cannot be compelled to provide broadband on a common carrier basis. . . . [T]he enormous regulatory scaffold that the Commission has built up around incumbent LEC xDSL offerings must be dismantled.").

[193] *See, e.g., In re Inquiry Concerning High-Speed Access to the Internet Over Cable and Other Facilities, Notice of Inquiry*, 15 FCC. Rcd. 19287, ¶ 13 & nn.25-29 (2000) (discussing several such attempts).

[194] *See MediaOne Group, Inc. v. County of Henrico, Va.*, 257 F.3d 356, 359, 363-64 (4th Cir. 2001) (holding that local "open access" laws are preempted by Section 621(b)(3)(D) of the Cable Communications Policy Act of 1984, as amended by the Telecommunications Act of 1996, 47 U.S.C. § 541(b)(3)(D)); *AT&T Corp. v. City of Portland*, 216 F.3d 871, 877-79 (9th Cir. 2000) (same).

Despite (or perhaps because of) the intensity of the controversy, however, the FCC thus far has assiduously avoided deciding whether or not cable modem service providers should be required to unbundle high-speed data transmission services from ISP services, or to provide transmission services to unaffiliated ISPs on nondiscriminatory terms and conditions. Indeed, for years, the FCC refused even to rule on the threshold regulatory question whether the transmission component of cable modem service is properly characterized as a "cable service," an "information service," a "telecommunications service," or some combination thereof.[197] Even while this threshold question remained unsettled, however, the Commission was forced to adjudicate several *ad hoc* demands for attachment of "cable open access" conditions that arose during proceedings in which the Commission's regulatory approval was needed for several large individual cable mergers to proceed. The results of these merger proceedings were inconsistent, however: "[t]he FCC rejected calls for open access in the AT&T/TCI and AT&T/Media One merger decisions, but both the FTC and the FCC imposed open access rules on the AOL/Time Warner merger."[198]

---

[195] *See MediaOne Group, Inc. v. County of Henrico, Va.*, 257 F.3d 356, 365 (4th Cir. 2001) (although local governments are preempted from requiring cable open access, "the merits of open access are not before us. . . . For the time being, therefore, we are content to leave these issues to the expertise of the FCC."); *AT&T Corp. v. City of Portland*, 216 F.3d 871, 879-880 (9th Cir. 2000) ("Thus far, the FCC has not subjected cable broadband to any regulation, including common carrier telecommunications regulation. . . . Congress has reposed the details of telecommunications policy in the FCC, and we will not impinge on its authority over these matters.").

[196] *Cf. GTE.Net LLC v. Cox Communications, Inc.*, 185 F.Supp.2d 1141 (S.D. Cal. 2002) (invoking doctrine of primary jurisdiction to defer decision to FCC on whether "cable open access" is mandated by Communications Act).

[197] *See, e.g.*, *National Cable & Telecommunications Ass'n, Inc. v. Gulf Power Co.*, 122 S. Ct. 782, 797-98 & n.4-6 (2002) (Thomas, J., concurring in part and dissenting in part) (surveying history of FCC's refusal to classify the service that is provided by cable transmission of data between customer's home and ISP's point-of-presence on the Internet) (citing FCC Orders, FCC *amicus* briefs in circuit court cases, and FCC petition for certiorari in *Gulf Power Co.* case); *see also* Rosemary Harold, **Cable-Based Internet Access: Exorcising the Ghosts of "Legacy" Regulation,** 28 N. Ky. L. Rev 721, 728 (2001) (noting the "court decisions spawned by the agency's deliberate inaction" in refusing to adopt a classification).

[198] James B. Speta, *A Common Carrier Approach to Internet Interconnection*, 54 Fed. Comm. L.J. 225, 234 (2002). *See also In re AOL Time Warner Inc.*, 16 FCC Rcd. 6547, ¶¶ 17-18 (2001) (requiring the merging parties, *inter alia*, "to afford access to Time Warner's cable plant to unaffiliated ISPs . . . and to hold separate Road Runner, [Time

Finally, on March 15, 2002, the FCC classified the transmission component of cable modem service as an "information service."[199] In so doing, the FCC did not immediately implement cable open access, nor did it disclaim any authority or intention to do so. Instead, by classifying every aspect of cable modem service as an "information service" the FCC effectively protected its own future discretion to adopt—or not to adopt—cable open access.[200] Had the Commission classified the transmission

---

Warner's affiliated] cable ISP, from AOL's ISP service until AOL Time Warner offers an unaffiliated ISP on all AOL Time Warner cable systems"). *But compare Transfer of Control of Licenses From Tele-Communications, Inc. To AT&T Corp.*, 14 FCC Rcd. 3160, ¶ 96 (1999) (refusing to impose cable modem "open access" as a condition of approving the AT&T/TCI merger); *Transfer of Control of Licenses From MediaOne Group, Inc. to AT&T Corp.*, 15 FCC Rcd. 9816, ¶¶ 126-28 (2000) (refusing to impose cable modem "open access" as a condition of approving the AT&T/MediaOne merger), *recon. denied*, 16 FCC Rcd. 5610 (2001), *compliance deadlines suspended*, 16 FCC Rcd. 5835 (2001), *recon. denied*, 16 FCC Rcd. 20587 (2001); *but see id*. ¶¶ 120-21 (noting that AT&T/MediaOne had committed to voluntarily open its cable modem platform to unaffiliated ISPs by June 2002, and to offer reasonably comparable access prices to unaffiliated ISPs).

[199] *See In re Inquiry Concerning High-Speed Access to the Internet Over Cable and Other Facilities, Declaratory Ruling & NPRM*, FCC 02-77, GEN Docket No. 00-185, 2002 WL 407567 (rel. Mar. 15, 2002). The Telecommunications Act of 1996 defines "information service" as "the offering of a capability for generating, acquiring, storing, transforming, processing, retrieving, utilizing, or making available information via telecommunications, and includes electronic publishing, but does not include any use of any such capability for the management, control, or operation of a telecommunications system or the management of a telecommunications service." 47 U.S.C. § 153(20); *see also High-Speed Access to the Internet Over Cable and Other Facilities, Declaratory Ruling & NPRM*, FCC 02-77, GEN Docket No. 00-185, 2002 WL 407567, at ¶ 34 n.139 (rel. Mar. 15, 2002) (tracing etymology of the "information services" concept). In 1999, the FCC had already classified the "ISP" component of cable modem service as an "information service." *See National Cable & Telecommunications Ass'n, Inc. v. Gulf Power Co.*, 122 S. Ct. 782, 796-97 n.4 (2002) (Thomas, J., concurring in part and dissenting in part) (citing *Deployment of Wireline Services Offering Advanced Telecommunications Capability*, 15 FCC Rcd. 385, 401 (1999), *vacated in part in other respects*, *WorldCom, Inc. v. FCC*, 246 F.3d 690 (D.C. Cir. 2001), *and otherwise aff'd*, *Association of Communications Enterprises v. FCC*, 253 F.3d 29 (D.C. Cir. 2001)). The "ISP" component of cable modem service is the component in which data is transferred from the provider's point-of-presence at the cable headend through a portal and onto the Internet. *See id.*

[200] Interestingly, some commentators in the cable modem rulemaking proceeding had expressly proposed that "[t]he Commission's decision of how to classify cable modem service . . . should be outcome-oriented: the Commission has substantial discretion in

(continued . . .)

component of cable modem service as a "telecommunications service" rather than an "information service," it might have been *required* to implementation cable open access, because telecommunications carriers are generally subject to common carrier regulation and unbundling requirements.[201] Conversely, classifying cable modem service as a "cable service" would likely have *prohibited* the FCC from implementing cable open access, while simultaneously empowering local cable franchising authorities to do so.[202]

In contrast, the Communications Act neither requires nor clearly prohibits the FCC from requiring unbundling of "information services."[203] Accordingly, in the same

---

how to classify the service and should thus consider the effect of its decision on competition in the broadband marketplace and the options it reserves for itself in the future." Comments of the Competition Policy Institute *filed In re Notice of Inquiry Concerning High-Speed Access to the Internet Over Cable and Other Facilities*, GEN Docket No. 00-185, at 1-2 (filed Dec. 1, 2000) (responding to NOI published at 15 FCC. Rcd. 19287 (2000)). Although the classification of cable modem services as "information services" did in fact maximize the Commission's reservation of power to exercise future discretion, Chairman Powell specifically disclaimed that the classification was the product of result-oriented decision-making. *See In re Inquiry Concerning High-Speed Access to the Internet Over Cable and Other Facilities, Declaratory Ruling & NPRM*, FCC 02-77, GEN Docket No. 00-185 (rel. Mar. 15, 2002) (Separate Statement of Chairman Powell) ("The Commission is not permitted to look at the consequences of different definitions and then choose the label that comports with its preferred regulatory treatment. That would be contrary to law. The Commission must apply the definition and then accept the regulatory regime that adheres to that classification and that which Congress chose when it adopted the statute.").

[201] *See Notice of Inquiry Concerning High-Speed Access to Internet*, 15 FCC Rcd. 19287, ¶ 20 (2000) (noting that the Communications Act "imposes a wide variety of obligations on telecommunications carriers, including requirements relating to interconnection, universal service contributions, disabilities access, and privacy of subscriber information."). On the other hand, "the FCC has broad authority to forbear from enforcing the telecommunications provisions if it determines that such action is unnecessary to prevent discrimination and protect consumers, and is consistent with the public interest." *AT&T Corp. v. City of Portland*, 216 F.3d 871, 878 (9th Cir. 2000) (citing 47 U.S.C. § 160(a)). Therefore, classifying the transmission component of cable modem service as a "telecommunications service" would likely only have established a rebuttable presumption in favor of cable open access.

[202] *See AT&T Corp. v. City of Portland*, 216 F.3d 871, 876-78 (9th Cir. 2000) (discussing the regulatory consequences of classifying cable modem service as a "cable service").

[203] *See In re Inquiry Concerning High-Speed Access to the Internet Over Cable and Other Facilities, Declaratory Ruling & NPRM*, FCC 02-77, GEN Docket No. 00-185, 2002 WL 407567, at ¶¶ 75-79 (rel. Mar. 15, 2002). *See also* Part VI.C, *infra* (discussing

order in which it adopted the classification, the FCC simultaneously launched a new rulemaking proceeding to consider the merits of "whether it is necessary or appropriate at this time to require that cable operators provide unaffiliated ISPs with the right to access cable modem service customers directly. . . ."[204]

## VI. Cable Open Access vs. Direct Access To INTELSAT in Comparative Perspective

The arguments for and against cable open access, discussed in Part V.B, above, center around four relatively uncontroversial policy goals. "These include: (1) encouraging deployment of facilities that provide advanced services to the widest possible spectrum of Americans; (2) encouraging competition among providers of broadband service; [and] (3) maintaining neutrality among the technologies that provide broadband Internet services."[205] While these policy goals, stated abstractly, are not controversial, the question of whether cable open access will contribute to their achievement has generated substantial controversy.[206] As discussed in Part V.A, above,

---

the FCC's statutory authority to require unbundling of information services).

[204] *In re Inquiry Concerning High-Speed Access to the Internet Over Cable and Other Facilities, Declaratory Ruling & NPRM*, FCC 02-77, GEN Docket No. 00-185, 2002 WL 407567, ¶ 72 (rel. Mar. 15, 2002), *petitions for review pending sub nom.*, *EarthLink, Inc. v. FCC*, Docket No. 02-1097 (D.C. Cir. filed Mar. 26, 1997). In this new rulemaking proceeding, the Commission continued to solicit reasons why the Commission might lack legal authority to implement cable open access. *See id.* ¶¶ 72, 79 (soliciting further comment on the scope of the FCC's statutory authority to implement direct access); *id.* ¶¶ 80-82 (soliciting comment on possible constitutional limitations on the FCC's authority to implement direct access). The Commission also made clear, however, that it is now ready to consider the substantive merits of implementing such a policy *See id.* ¶ 73 (setting forth substantive considerations which the FCC seeks comment on).

[205] Comments of the Competition Policy Institute *filed In re Notice of Inquiry Concerning High-Speed Access to the Internet Over Cable and Other Facilities*, GEN Docket No. 00-185, at 2 (filed Dec. 1, 2000) (responding to NOI published at 15 F.C.C. Rcd. 19287 (2000)). The Competition Policy Institute also notes a fourth policy goal of "removing regulation where the public interest is served by that action." *Id.* A general presumption favoring deregulation would, of course, weigh against regulatory imposition of either cable open access *or* INTELSAT direct access. However, because this policy goal would not appear to provide a basis for comparison between cable open access and INTELSAT direct access, it is not discussed herein.

[206] More than 300 public Comments and Reply Comments were filed in response to the FCC's Notice of Inquiry concerning cable open access. *See In re Notice of Inquiry Concerning High-Speed Access to the Internet Over Cable and Other Facilities*, 15 FCC Rcd. 19287, GEN Docket No. 00-185 (rel. Sept. 28, 2000).

the debate over open access to INTELSAT has revolved around substantially the same policy issues, and has generated substantially similar controversies concerning the efficacy of access regulation in advancing these goals. This Part compares the efficacy of "cable open access" with the efficacy of "INTELSAT direct access" in achieving each of these goals.

### A. Encouraging Deployment of Facilities.

Since 1934, it has been "the policy of the United States to encourage the provision of new technologies and services to the public."[207] In 1996, Congress reaffirmed this policy when it directed the FCC (and state Public Utilities Commissions) to "encourage the deployment on a reasonable and timely basis of advanced telecommunications capability to all Americans . . . by utilizing, in a manner consistent with the public interest, convenience, and necessity, price cap regulation, regulatory forbearance, measures that promote competition in the local telecommunications market, or other regulating methods that remove barriers to infrastructure investment."[208] Accordingly, the Commission has recently characterized its goal of "promot[ing] widespread and rapid deployment of high-speed services" as one of its prime statutory objectives.[209]

Similarly, encouraging the rapid deployment of new communications facilities was also one of the principle goals of the Communications Act of 1962.[210] Indeed, the

---

[207] 47 U.S.C. § 157(a).

[208] Telecommunications Act of 1996 §706(a), Pub. L. No. 104-104 § 706(a), 110 Stat. 56, 153 (1996), *codified in notes following* 47 U.S.C. §157 (2000). The 1996 Act defines "advanced telecommunications capability" to mean *all* "high-speed, switched, broadband telecommunications capability that enables users to originate and receive high-quality voice, data, graphics, and video telecommunications using any technology." *Id.* § 706(c)(1), *codified in notes following* 47 U.S.C. §157 (2000). The FCC implemented Section 706 in *In re Inquiry Concerning the Deployment of Advanced Telecommunications Capability to All Americans in a Reasonable and Timely Fashion, Second Report*, 15 FCC Rcd. 20913 (2000).

[209] *In re Notice of Inquiry Concerning High-Speed Access to the Internet Over Cable and Other Facilities, Notice of Inquiry*, 15 F.C.C. Rcd. 19287, ¶ 2 (2000); *accord id.* ¶ 12 (promising to "continue to monitor broadband deployment closely to see whether there are developments that could affect our goal of encouraging deployment of broadband capabilities pursuant to the requirements of section 706.") (quoting *First 706 Report,* 14 FCC Rcd at 2449 ¶ 101).

[210] *See, e.g.*, 47 U.S.C. § 701(a) ("is the policy of the United States to establish, in conjunction and in cooperation with other countries, *as expeditiously as practicable* a commercial communications satellite system . . . which will serve the communication needs of the United States and other countries, and which will contribute to world peace

(continued . . .)

Satellite Act expressly mandated that "[t]he new and expanded [satellite] telecommunication services *are to be made available as promptly as possible* and are to be extended to provide global coverage *at the earliest practicable date*."[211]

Proponents of "cable open access" have asserted that such regulatory unbundling requirements would best serve Congress's goal of encouraging rapid deployment of facilities. Specifically, proponents warn that prohibitive entry costs may indefinitely hinder new entrants from building "last-mile" cable connections to homes that would be capable of providing end-to-end facilities-based competition against incumbent high-speed cable ISPs.[212] The cost to a potential new entrant of constructing Internet backbone and ISP facilities, however, may be less prohibitive than the cost of constructing "last-mile" connections to the home.[213] Accordingly, based on their prediction that competition in the "last mile" will not come soon under any regulatory regime,[214] proponents have asserted that an open access policy would create incentives for competitors unequipped to deploy their own "last mile" facilities to, at least, deploy additional backbone and ISP facilities that they would otherwise have no reason to

---

and understanding.") (emphasis added).

[211]    *Id.* § 701(b) (emphasis added).

[212]    Cable open access proponents, for example, have asserted that "[s]ince cable currently controls over 80% of the high-speed Internet access market, there is effectively no significant competition in the broadband Internet access market." Comments of the Open Net Coalition *filed In re Notice of Inquiry Concerning High-Speed Access to the Internet Over Cable and Other Facilities*, GEN Docket No. 00-185, at 2 (filed Dec. 1, 2000), *available online at* <http://www.opennetcoalition.org/resources/NOIfiling.pdf>.

[213]    *See, e.g.*, Rob Frieden, *Does A Hierarchical Internet Necessitate Multilateral Intervention?*, 26 N.C. J. Int'l L. & Com. Reg. 361, 370 (2001) ("The facilities-based, long-haul telecommunications transmission marketplace has such substantial market entry and operational costs that relatively few operators can efficiently and effectively enter and remain in the market. This . . . contrasts with the comparatively low costs and low barriers to market entry in *reselling* the long-haul services of a Tier-1 [facilities-based] ISP") (emphasis added and footnote omitted).

[214]    *See* Comments of the Open Net Coalition *filed In re Notice of Inquiry Concerning High-Speed Access to the Internet Over Cable and Other Facilities*, GEN Docket No. 00-185, at 4 (filed Dec. 1, 2000), *available online at* <http://www.opennetcoalition.org/resources/NOIfiling.pdf> ("As a practical matter, only two methods of two-way high-speed Internet access . . . are presently available for most Internet subscribers in the United States—cable Internet access and DSL. DSL, however, has technical limitations which make it unavailable to many Americans.") (footnote omitted).

deploy.[215] Finally, proponents have also asserted that as these unaffiliated backbone and ISP providers began to generate business and revenues, they would eventually grow to the point where they would find it feasible (and desirable) to build their own competitive "last-mile" facilities.

Similarly, "INTELSAT direct access" proponents also asserted that regulatory unbundling requirements would best encourage rapid deployment of facilities. Like construction of a citywide cable system, construction and launch of a geostationary communications satellite can be prohibitively expensive: the cost of constructing and launching a single geostationary satellite normally exceeds $500,000,000,[216] and a substantial proportion of satellite launch attempts either explode on the launching pad or

---

[215] *See id.* at 2, 19-20 (filed Dec. 1, 2000) (asserting that "a failure to implement cable open access "will leave millions of high-speed cable Internet consumers with no choice but to accept and pay for ISPs that are owned or affiliated with the cable operator," and will thereby "frustrate, delay, or simply deny" the rapid deployment of high-speed cable Internet services).

It should be noted that there is some controversy regarding whether deployment of additional non-last-mile ISP facilities *does* add value to the Internet. *See Technological and Regulatory Factors Affecting Consumer Choice of Internet Providers*, United States General Accounting Office Report to the Sen. Judiciary Comm., Subcomm. on Antitrust, Business Rights, and Competition, GAO-01-93, at 30 (Oct. 2000) ("The experts and industry officials we interviewed differed over whether a reduction in ISP choice—if it occurs—constitutes a public policy concern. Some experts felt that . . . a reduction of consumer choice at the ISP layer is not a concern as long as there is adequate competition among companies providing physical transport to the Internet"). For articulations of this view, *see. e.g.*, Rob Frieden, *Does A Hierarchical Internet Necessitate Multilateral Intervention?*, 26 N.C. J. Int'l L. & Com. Reg. 361, 369 (2001) ("In general, a healthy and efficiently operating Internet industry can exist even under a hierarchical structure coupled with a limited number of Tier-1 ISPs.") (footnote omitted); *accord* Comments of National Cable Television Association *filed In re Notice of Inquiry Concerning High-Speed Access to the Internet Over Cable and Other Facilities*, GEN Docket No. 00-185, at 53 (filed Dec. 1, 2000) (same); Comments of Cox Communications *filed In re Notice of Inquiry Concerning High-Speed Access to the Internet Over Cable and Other Facilities*, GEN Docket No. 00-185, at 31 (filed Dec. 1, 2000) (same). For present purposes, however, this Paper will assume that the FCC would conclude that additional deployment of non-last-mile ISP services will yield a marginal public benefit.

[216] *See, e.g.*, Andy Pasztor, *Hughes, Lockheed Satellite Projects Lack Sponsors*, Wall St. J., June 4, 2001, at A3, 2001 WL-WSJ 2865370 (projecting a cost between $4.4 billion and $5.2 billion for construction and launch of eight to twelve proposed geostationary satellites).

otherwise fail to achieve orbit successfully.[217] The cost of constructing new satellite earth stations, and new microwave and fiber optic links to connect earth stations with end users, however, may not be as prohibitive as the cost of constructing and launching a geostationary space station. Accordingly—and analogously with the deployment-based arguments advanced by proponents of cable open access—proponents asserted that INTELSAT direct access would expedite the deployment of satellite earth stations and terrestrial microwave and fiber optic link facilities, by making it possible for unaffiliated carriers to compete against COMSAT for business delivering INTELSAT communications services to end users.[218] Finally, direct access proponents—again, like cable open access proponents—also asserted that the revenues generated by unaffiliated carriers operating under direct access would eventually make it possible for those carriers to launch their own separate geostationary satellite systems.

Opponents of both "cable open access" and "INTELSAT direct access," however, have taken issue with the proponents' assumption that end-to-end facilities-based competition would (or will) never arrive. Although cable systems have not been overbuilt by competing cable systems in very many localities, cable ISPs *have* begun to face intermodal facilities-based competition from competing technologies such as DSL that are capable of providing similar high-speed transmission services.[219] Similarly, INTELSAT—despite its history as a protected monopoly—*has*, in recent years, faced facilities-based competition from both competing geostationary satellite systems and from competing technologies (*e.g.*, transoceanic submarine cables) capable of providing similar international telecommunications transmission services.[220] Because end-to-end facilities-based competition *is* economically feasible, opponents assert, regulatory access requirements will actually *retard* deployment of new facilities, by enabling putative competitors to obtain market entry without making any risky or expensive investments in competing end-to-end facilities.[221] Moreover, opponents assert that access requirements "inhibit the investment necessary to the continued deployment of new technologies and rapid deployment of broadband capabilities. . . [by] seriously undermin[ing] the . . . incentives to make such [investments]."[222] A leading opponent has stated that under a "cable open access" regime:

> There would be reduced reasons for cable operators to take the substantial risks associated with the deployment of new facilities and services if, from the first day, they were burdened by onerous and burdensome government regulation that forced them to make the broadband capabilities of their cable plant available to competing Internet service providers that have chosen not to take those risks. . . . The prospect of ill-defined and far-reaching regulation would . . . diminish the ability of corporate entities to

---

[217] *See generally* House of Representatives Select Committee, Report on U.S. National Security and Military/Commercial Concerns with the People's Republic of China, H.R. Rep. No. 105-851, ch. 5-8 (1999), *available online at* <http://www.house.gov/coxreport/cont/gncont.html> (discussing several satellite launch failures).

[218] *See Direct Access To The INTELSAT System*, 14 FCC Rcd 15703, ¶ 16 (1999) (citing carrier comments) . *See also id.* ¶ 42 (concluding that while "direct access . . . does not add another facilities-based competitor, the additional choice, flexibility, and cost savings made available by direct access to U.S. customers in use of an existing facilities-based provider—INTELSAT—would result in increased competition."). Indeed, in the 1980s, some proponents asserted that direct access would yield incentives for unaffiliated carriers to invest in INTELSAT itself, thereby enhancing deployment of additional INTELSAT satellite facilities. *1982 Direct Access NOI*, 90 F.C.C. 2d 1446, ¶ 8 (1982); *see also 1984 Direct Access Order*, 97 FCC 2d 296, ¶¶ 14-15 (1984). An analogue to this argument in today's cable open access debate would be an assertion that cable open access would encourage unaffiliated ISPs (*e.g.*, EarthLink) to invest in facilities-based cable system operators (*e.g.*, AT&T/TCI), and that such investment would enable the cable operators to deploy additional last mile facilities. No such argument, however, appears to have been advanced by any cable open access proponent to date.

[219] *See* Part II.B, *supra*.

---

[220] *See* Part II.C, *supra*.

[221] *See, e.g.*, Comments of AT&T *filed In re Notice of Inquiry Concerning High-Speed Access to the Internet Over Cable and Other Facilities*, GEN Docket No. 00-185, at 40-41 (filed Dec. 1, 2000) ("By forbearing from imposing 'open access' regulations on cable operators, the Commission has fostered an environment that encourages investment not only in cable, but also in the alternative broadband technologies, wireless, satellite, and DSL") (quoting FCC Cable Services Bureau, *Broadband Today*, Rep. No. CS 99-14, at 49 (Oct. 1999), *available online at* <http://www.fcc.gov/Bureaus/Cable/Reports/broadbandtoday.pdf>); *accord* Comments of COMSAT Corp. *filed In re Direct Access To The INTELSAT System*, IB Docket No. 98-192, at 53 (filed Dec. 23, 1998) (noting that a massive deployment of separate international satellite systems and transoceanic submarine cables had "all taken place in the absence of direct access. And it is not at all clear how substituting INTELSAT in the U.S. for COMSAT will engender more competition. . . ."); *cf. id.* at 61 (asserting that direct access would distort the market and discourage deployment of competing facilities by causing INTELSAT space segment capacity to be priced below-cost).

[222] Comments of AT&T *filed In re Notice of Inquiry Concerning High-Speed Access to the Internet Over Cable and Other Facilities*, GEN Docket No. 00-185, at 68 (filed Dec. 1, 2000); *accord id.* at 66 (asserting that a cable open access requirement would "dramatically slow deployment of broadband access, deter investment, stall development of new services and technologies, [and] discourage innovative business models. . . .").

plan new buildouts, [and also] would effectively kill the public
equity market for financing.²²³

The *absolute* merits of deployment-based arguments for both "cable open access" and "INTELSAT direct access" are subject to reasonable debate. In comparative perspective, however, the deployment-based case for "cable open access" appears substantially stronger than the deployment-based case for "INTELSAT direct access." In 1999, when the FCC adopted a policy of direct access to INTELSAT, INTELSAT satellites were already subject to substantial facilities-based competition on every major telecommunications route to or from the United States. During that year, INTELSAT owned and operated only 15 of the approximately 75 geosynchronous commercial communications satellites that were then capable of serving the United States.²²⁴ Moreover, by 1999, transoceanic submarine fiber optic cables already delivered about three times the amount of U.S. international transmission circuits delivered by all satellite providers—including INTELSAT—combined.²²⁵ Because of this massive deployment of competitive facilities during the decade before "direct access" to INTELSAT was mandated, INTELSAT's share in the international telecommunications transmission market dropped from nearly 100% in the mid-1980s, to less than half by the late 1990s.²²⁶ Today, even while remaining subject to the unique "direct access" requirement, the recently privatized Intelsat is now estimated to serve only 14% of the market for international *satellite* services,[227] which is, itself, only one-third of the overall market for international transmission services.[228]

---

²²³  *Id.* at 68-69 (internal punctuation marks and citations omitted).

²²⁴  *See Phillips Satellite Industry Directory*, at 17-234, 279-413 (21st ed. 1999) (setting forth complete information about each of these satellites and their operators). In 1999, a total of nearly 200 total geosynchronous commercial communications satellites were orbiting the earth. *Id.* Of these 200 satellites, INTELSAT owned and operated 19. *Id.* Four of INTELSAT's 19 satellites, however, were incapable of serving the United States because they were located above the Indian Ocean. *Id.*

²²⁵  *See* Part II.C, *supra*. *See also, e.g. In re COMSAT Corp. Reclassification as a Non-Dominant Carrier*, 13 FCC Rcd. 14083, ¶¶ 11, 19, 32-39 (1998) ("*COMSAT Non-Dominant Order*"), *modified on recon.*, 14 FCC Rcd. 3065 (1999) (characterizing satellites and submarine cables as fungible commodities serving the markets for switched voice, private line, and video services, and noting that cables compete effectively against INTELSAT satellites on every major international telecommunications route to or from the United States); *id.* at 14131 (less than 19,000 satellite circuits carried U.S.-international traffic, compared to more than 57,000 submarine cable circuits).

²²⁶  *See id.* at 14121, 14131, 14134-35 (INTELSAT's share of international switched voice and private line traffic to and from the United States decreased from an average of 70% in 1988 to less than 21% in 1996; its share of the U.S. international video transmission market dropped from 80% in 1994 to less than 45% in 1996).

Like INTELSAT, cable ISPs also face substantial facilities-based competition, at least from intermodal competitors.[229] But unlike INTELSAT, cable ISPs have not yet surrendered the lion's share of the their market share to their competitors, facilities-based or otherwise. Rather, as of August 2001, cable modems continued to serve 68% of the market for residential high-speed Internet access,[230] and "their upgraded networks are far more ubiquitous than any competing networks."[231] Moreover, although other high-speed transmission technologies (especially DSL) are currently gaining market share, the FCC has nonetheless projected that cable will continue to lead the broadband market until at least 2007.[232]

The theory that a government-mandated access requirement will spur—rather than hinder—deployment of new facilities can apply only if some structural impediment (such as conditions of "natural monopoly") is impairing the natural development of end-to-end facilities-based competition.²³³ As discussed above, both the residential high-speed Internet transmission market and the international telecommunications transmission market seem to be moving inexorably from conditions of *de facto* monopoly to conditions of increasing competition. However, the international telecommunications transmission market (which is now fully competitive) has moved much further down this path than has the residential high-speed Internet transmission market (which cable ISPs still dominate). Accordingly, the deployment-based justification for "cable open access" is comparatively stronger than the deployment-based justification for "direct access to INTELSAT."

---

[227] *Players Ready For Further Consolidation*, INTERSPACE, April 11, 2001, 2001 WL 10292682. INTELSAT's share in the U.S.-international satellite communications market is now third to its intramodal competitors Hughes/PanAmSat Corp. (36 percent) and General Electric/GE Americom (29 percent). *Id.*

[228] *COMSAT Non-Dominant* Order, 13 FCC Rcd. At 14131.

[229] *See* Part II.B, *supra*. *See also, e.g.*, *Transfer of Control of Licenses and Section 214 Authorizations from MediaOne Group, Inc. to AT&T Corp.*, 15 FCC Rcd. 9816, at ¶¶ 116-23 (2000) ("*AT&T/MediaOne Order*")) (declining to impose an open access condition because of, *inter alia*, "the increasingly rapid deployment of alternative high-speed Internet platforms, especially xDSL").

[230] *In re Inquiry Concerning High-Speed Access to the Internet Over Cable and Other Facilities, Declaratory Ruling & NPRM*, FCC 02-77, GEN Docket No. 00-185, 2002 WL 407567, ¶ 9 (rel. Mar. 15, 2002); *accord* U.S. Department of Commerce, National Telecommunications and Information Administration, *A Nation Online: How Americans Are Expanding Their Use of the Internet* 35 (Feb. 2002), available online at <http://www.ntia.doc.gov/ntiahome/dn/anationonline2.pdf>.

[231] Comments of SBC Communications Inc. & BellSouth Corp. *filed In re Notice of Inquiry Concerning High-Speed Access to the Internet Over Cable and Other Facilities*, GEN Docket No. 00-185, at ii (filed Dec. 1, 2000); *accord id.* at 5 & nn.12-13 (citing several market share studies).

[232] *See* FCC Cable Services Bureau, *Broadband Today*, Rep. No. CS 99-14, at 27 & App. B, Chart 2 (Oct. 1999), *available online at* <http://www.fcc.gov/Bureaus/Cable/Reports/broadbandtoday.pdf>).

### B. Encouraging Competition Among Service Providers.

Whether "cable open access" or "direct access to INTELSAT" will enhance or diminish competition in their respective markets are questions closely related to the issues, discussed in Part VI.A, *supra*, of whether such mandated access requirements will encourage or discourage rapid deployment of competitive facilities. Increased competition, of course, normally brings additional benefits to consumers, beyond merely encouraging investment in the deployment of new facilities. Among the benefits of competition that Congress has generally directed the FCC to promote are enhanced consumer choice, lower consumer prices, better quality of service, and innovative service offerings.²³⁴ The debate over whether government-mandated access requirements will help achieve such benefits, however, parallels in many respects the debate over whether such requirements will help achieve the related goal of encouraging rapid deployment of new telecommunications facilities.

Cable open access proponents, for example, have asserted that since cable currently controls a dominant share of the residential high-speed Internet access market, "there is effectively no significant competition in [that] market."²³⁵ Accordingly, these

---

²³³ *See* note **[__]**, *supra*.

²³⁴ *See, e.g.*, *AOL/Time Warner Merger, Mem. Opinion & Order*, 23 Comm. Reg. (P&F) 157, ¶ 59 & n.169 (Jan. 22, 2001) ("[I]n adopting the 1996 Act, Congress established a clear national policy to 'promote the continued development of the Internet' and 'to preserve the vibrant and competitive free market that presently exists for the Internet and other interactive computer services unfettered by Federal or State regulation.'") (citing 47 U.S.C. §§ 230(b)(1)-(2)); *see also* Comments of the Open Net Coalition *filed In re Notice of Inquiry Concerning High-Speed Access to the Internet Over Cable and Other Facilities*, GEN Docket No. 00-185, at 1-2 (filed Dec. 1, 2000), *available online at* <http://www.opennetcoalition.org/resources/NOIfiling.pdf> (specifically identifying the benefits of competition as including consumer choice, lower prices, better quality of service, and innovative service offerings).

²³⁵ Comments of the Open Net Coalition at 2. *See also Technological and Regulatory Factors Affecting Consumer Choice of Internet Providers*, United States General Accounting Office Report to the Sen. Judiciary Comm., Subcomm. on Antitrust,

(continued . . .)

proponents assert that a mandatory cable open access requirement would enhance consumer choice by "ensur[ing] that the next generation of Internet access develops with the open architecture and accompanying vibrant competition that has characterized the development of a competitive Internet to date."[236] On this theory, a failure to implement cable open access "will leave millions of high-speed cable Internet consumers with no choice but to accept and pay for ISPs that are owned or affiliated with the cable operator."[237] If so, access proponents assert, then cable operators will have a means of forcing consumers to pay for additional unwanted services in order to obtain the package of high-speed Internet services they desire.[238] Alternatively, proponents assert that cable-affiliated ISPs may also have incentives to restrict consumers' access to Internet content—especially "streaming video"—that threatens their core cable television business.[239] Finally, proponents assert that without cable open access, cable ISPs will extract monopoly rents in the form of excessive rates charged to their residential customers.[240]

---

Business Rights, and Competition, GAO-01-93, at 44 (Oct. 2000) ("*GAO Consumer Choice Study*") ("[I]f broadband is a distinct market, cable firms do currently hold a leading position in that market.").

[236] Comments of the Open Net Coalition at 19; *see also GAO Consumer Choice Study*, GAO-01-93, at 23-24 (Oct. 2000) (noting that "the common carrier status of telephone companies, which requires that they provide nondiscriminatory service at just and reasonable rates, worked to give [narrowband] ISPs easy access to consumers through the telephone network").

[237] Comments of the Open Net Coalition at 19-20.

[238] *Id.* at 2; *see also id.* at 6 ("Because many ISPs have become 'content aggregators' offering varied content and diverse applications—in addition to basic 'on-ramp' capability, preserving choice in the ISP marketplace is key to allowing consumers to 'vote with their feet' and switch ISPs if they do not like the content and/or applications of the ISP affiliated with their cable modem service provider.") (footnote and citation omitted) (quoting *GAO Consumer Choice Study*, GAO-01-93, at 30 (Oct. 2000)). *Cf.* Comments of the Competitive Access Coalition *filed In re Notice of Inquiry Concerning High-Speed Access to the Internet Over Cable and Other Facilities*, GEN Docket No. 00-185, at 5 (filed Dec. 1, 2000) (characterizing the now-routine bundling of cable high-speed Internet transmission with affiliated cable ISP service as an "anticompetitive tying agreement[]").

[239] *See id.* at 8 (alleging that a cable ISP's recent decision to limit video streaming to ten minutes "likely resulted from a business decision by cable operators to limit competition to the proprietary cable television service also owned by the owners of cable modem service, thereby impeding the development of this important new technology by limiting its markets.").

[240] *See, e.g.*, Comments of the Competitive Access Coalition *filed In re Notice of*

(continued . . . )

Similarly, "INTELSAT direct access" proponents also asserted that regulatory unbundling requirements would best encourage such benefits of competition as enhanced consumer choice, lower consumer prices, better quality of service, and innovative service offerings. Specifically, such proponents "maintain[ed] that permitting direct access [would] promote competition and result in: (1) greater customer choice due to the availability of competitive alternatives for accessing INTELSAT (where INTELSAT is their system of choice); (2) opportunity for substantial cost savings as a result of competition for accessing INTELSAT, resulting in reduced end user prices; (3) greater customer control over service provision (involving service quality, performance costs, connectivity and redundancy); and (4) efficiencies in system planning and set up of circuits."[241]

As with the deployment-based arguments discussed in Part IV.A, *supra*, however, the question of whether mandated access requirements would yield the desired additional competitive benefits turns largely on the question of whether access requirements will spur increased competition or, conversely, hinder it. Seemingly, every identified benefit of competition would be maximized by the development of genuine facilities-based competition, as opposed to purely retail competition between resellers of transmission capacity provided using the same underlying facilities.[242] For this reason, mandatory unbundling requirements are desirably only in communications markets that enjoy no realistic near-term prospect for the development of facilities-based competition, or, alternatively, in markets where unbundling requirements would not reasonably be expected to create a disincentive to such development. Conversely, in markets where facilities-based competition might be expected to develop, unbundling requirements should not be adopted if they would create a disincentive to such development.

Like the deployment-based arguments, the *absolute* merits of competition-based arguments for both "cable open access" and "INTELSAT direct access" are subject to reasonable debate. In comparative perspective, however, the competition-based case for "cable open access" again appears substantially stronger than the competition-based case

---

*Inquiry Concerning High-Speed Access to the Internet Over Cable and Other Facilities*, GEN Docket No. 00-185, at 26 (filed Dec. 1, 2000) ("Any company, in the legitimate pursuit of its self-interest, will seek to exploit its control over a scarce resource. . . . It is natural for companies that control access to a connection point between producers and consumers to adopt strategies designed to maximize the profit potential of that control.").

[241] *See Direct Access To The INTELSAT System*, 14 FCC Rcd 15703, ¶ 16 (1999) (citing Comments filed by direct access proponents).

[242] Under cable open access, unaffiliated ISPs would resell high-speed transmission capacity provided by a single monopoly cable system operator. Under INTELSAT direct access, retail international carriers would resell satellite space segment transmission capacity provided by INTELSAT.

for "INTELSAT direct access." In 1999, when it implemented direct access to INTELSAT, the FCC explained the competitive benefits of direct access as follows:

> While . . . direct access . . . does not add another facilities-based competitor, the additional choice, flexibility, and cost savings made available by direct access to U.S. customers in use of an existing facilities-based provider—INTELSAT—would result in increased competition. . . . [D]irect access would place competitive pressures on other satellite operators in terms of service, price, and quality. In addition, it would place competitive pressures on Comsat, particularly with respect to services for which Comsat has a markup substantially higher than INTELSAT IUC [wholesale] rates.[243]

Like direct access to INTELSAT, cable open access would not add another facilities-based competitor. And like direct access to INTELSAT, cable open access would make available to U.S. customers the use of facilities that belong to an existing facilities-based provider—the incumbent cable operator's "last mile" connection—to obtain services from unaffiliated service providers (here, ISPs) that compete against the facilities-based provider's own affiliate. In the same way that direct access to INTELSAT "would place competitive pressures on" the affiliate (COMSAT) of the facilities-based provider (INTELSAT), cable open access "would place competitive pressures on" the affiliate (cable ISP) of the facilities-based provider (cable system operator).[244] In both cases, these "competitive pressures: would be particularly acute "with respect to services for which [the incumbent's retail affiliate] has a markup substantially higher than" its wholesale cost of obtaining transmission capacity from its facilities-based affiliated provider.[245]

There are, however, several competitive differences between direct access to INTELSAT and cable open access. First, while many (perhaps most) customers use their ISP "only [as] a simple 'on-ramp' to the Internet," an increasing number of customers select their ISP based on the individual ISP's "content and applications," which may include proprietary "search engines," "content aggregators," or a particular e-mail address domain.[246] Indeed, the most popular dial-up ISP, America Online, now offers a popular "'bring-your-own-access' plan providing unlimited access to thousands of unique AOL features . . . for individuals who already have an Internet connection or access through the work or school environment."[247] Because at least some customers are willing to pay for AOL's content and applications even while obtaining their basic "'on-ramp' to the Internet" elsewhere, it would appear that competition based on service offerings is possible between ISPs who provide identical transmission services.

---

[243] *Direct Access To The INTELSAT System*, 14 FCC Rcd 15703, ¶ 42 (1999).

[244] *Cf. id.*

[245] *Id.*

[246] *GAO Consumer Choice Study*, GAO-01-93, at 30 (Oct. 2000).

In the market for INTELSAT international space segment transmission capacity, in contrast, such competition is unlikely. INTELSAT space segment capacity is furnished in the form of "raw" transmission capacity to international common carriers (*e.g.*, AT&T, WorldCom, Sprint) and other users (*e.g.*, television networks, ISPs) who use the "raw" capacity as an input, to provide a component of the end-to-end telecommunications services that these entities produce. Accordingly, regardless of whether "raw" INTELSAT capacity is furnished by COMSAT, INTELSAT, or an unaffiliated "direct access" provider, the transmission capacity itself is identical. While INTELSAT satellites are subject to vigorous facilities-based competition based on price, availability, and quality of service, competing "direct access" providers of "raw" INTELSAT capacity have no meaningful way to compete with one another by "innovating" in the nature of their service offerings.

Cable open access is also more likely than INTELSAT direct access to result in meaningful price competition at the retail level. Currently, there are substantial variations in price between competing narrowband "dial-up" ISPs, all of whom normally must rely on the same local telephone company to transmit information between their own points-of-presence and their subscribers' homes. A number of advertiser-supported ISPs currently offer dial-up Internet service cost-free to the user.[248] Many no-frills ISPs offer unlimited monthly 56K dial-up Internet service for $9.95 per month or less.[249] On the other hand, AOL's "standard plan providing access to AOL and the Internet" currently retails for $23.90 per month.[250] This significant variation in the price charged by ISPs for traditional dial-up Internet access indicates that price competition between "pure" ISPs who do not provide "last-mile" facilities is possible, and may in fact exert downward pressure on ISP prices.

---

[247] AOL Anywhere Pricing Plans Web Page, <http://www.aol.com/info/pricing.html>.

[248] *See* Freedomlist Free ISP Web Page, <http://www.freedomlist.com/find.php3?country=166&st=1> (listing 12 U.S.-based ISPs that offer dial-up Internet access free of charge to the user).

[249] *See* Freedomlist Cheap ISP Web Page, <http://www.freedomlist.com/find.php3?country=166&st=3> (listing 124 U.S.-based ISPs that offer unlimited monthly 56K dial-up Internet access for less than $10 per month).

[250] AOL Anywhere Pricing Plans Web Page, <http://www.aol.com/info/pricing.html>.

In contrast, the INTELSAT wholesale IUC rate constitutes only an extremely minimal component of the retail cost of an international phone call. Thus, direct access to INTELSAT holds far less potential than cable open access to provide cost savings to end users. In 1984, the FCC declined to adopt direct access when, *inter alia*, it concluded that while "direct access might help trim the cost to U.S. users of INTELSAT space segment, it is doubtful that any savings in this area could exceed a few percentage points of the total cost to U.S. users of a communications channel."[251] Throughout the 1980s and 1990s, INTELSAT's wholesale IUC rate continued to drop, as technological advances enabled each new satellite to carry many more simultaneous calls than its predecessor. By 1999, when the FCC implemented direct access, an economic study conducted by the Harvard-affiliated Brattle Group estimated that the INTELSAT IUC rate constituted only 1.3% of the price of the average international phone call, meaning that any consumer "savings [from direct access] would amount to only 1.3% of the total end user charges, even if INTELSAT services were provided *free*."[252] The same study also concluded that even this potential 1.3% savings was highly likely to be appropriated by the U.S. and foreign international carriers who initiated and terminated the call, rather than passed through to the consumer.[253]

For these reasons, the competition-based justification for "cable open access" is comparatively stronger than the competition-based justification for "direct access to INTELSAT."

### C. Maintaining Regulatory Parity Among Technologies That Provide the Same (or Fungible) Services.

Historically, different types of communications facilities have been subject to different FCC regulations.[254] For much of the twentieth century, the FCC's atomized

---

[251] *Regulatory Policies Concerning Direct Access to INTELSAT Space Segment for the U.S. International Service Carriers,* 97 F.C.C. 2d 296, ¶ 49 (1984) ("*1984 Direct Access Order*"), aff'd, *Western Union Int'l, Inc. v. FCC*, 804 F.2d 1280 (D.C. Cir. 1986). *See also id.* ¶ 67 ("We have not been presented with any evidence to show that the alleged savings to be realized from these [direct access] proposals, assuming, *arguendo*, that such are passed-through dollar-for-dollar by carriers to end-users, would exceed more than a few percentage points of the total end-user charge.").

[252] Comments of COMSAT Corp. *filed in Direct Access To The INTELSAT System*, IB Docket No. 98-192, at 73 n.200 (filed Dec. 23, 1998) (emphasis in original).

[253] *Id.* at 74 & nn.201-02.

[254] *See* Part III, *supra* (discussing separate regulatory regimes that were employed to regulate cable, telephony, and satellite); *see generally* Rosemary Harold, **Cable-Based Internet Access: Exorcising the Ghosts of "Legacy" Regulation,** 28 N. Ky. L. Rev 721

(continued . . .)

regulatory approach made sense: cable television was not the same service as telephony, so why should it be subject to the same FCC regulations?[255] Indeed, even as technological advances began to facilitate the development of intermodal competition in telecommunications, the FCC sometimes continued to defend the propriety of certain legacy-based regulatory paradigms disparities which effectively imposed wildly disparate regulatory obligations on head-to-head competitors.[256]

By the mid-1990s, however, the FCC regularly asserted that when competing technological platforms are capable of providing fungible services, the public interest is normally best served by regulatory neutrality.[257] As one commentator explained, "[c]alls for fundamental equity and fairness of treatment are American maxims; in regulatory circles, arguments for 'leveling the playing field' exercise a great pull over time,

---

(2001) (discussing legacy of platform-specific regulation).

[255] Today, the FCC's Rules continue to be organized on a facility-specific basis. *See, e.g.*, 47 C.F.R. Part 25 (setting forth rules governing communications satellites); 47 C.F.R. Parts 42-69 (setting forth rules governing wireline common carriers); 47 C.F.R. Part 73 (setting forth rules governing broadcast radio and television); 47 C.F.R. Part 76 (setting forth rules governing cable TV).

[256] *See, e.g., Transfer of Control of McCaw Cellular Communications, Inc. to AT&T, Order on Recon.*, 10 FCC Rcd. 11786, ¶ 9 (1995) (asserting that the 1934 Communications Act requires the FCC "to focus on competition that benefits the public interest, not on equalizing competition among competitors. . . . [T]he Communications Act does not require parity between competitors as a general principle.") (footnotes and citations omitted); *see also MCI Telecommunications Corp.*, 68 F.C.C. 2d 1553, __ & n.16 (1978) (Washburn, Comm'r, concurring) ("The touchstone of regulation should be rooted in the public's interest, not in some notion of regulatory parity. This is especially true where the [regulated] parties are . . . dissimilarly situated. . . . It makes little sense to fashion the same cage for a canary as for a gorilla."), *aff'd sub nom., Lincoln Tel. & Tel. Co. v. FCC*, 659 F. 2d 1092 (D.C. Cir. 1981).

[257] *See, e.g., Waiver of the Commission's Rules Regulating Rates for Cable Services*, 11 FCC Rcd. 1179, ¶ 25 (1995) (noting that the public interest would be served by "establishing some measure of regulatory parity between the cable operators and [video dial tone] programmers" who provide similar services); *Equal Access and Interconnection Obligations Pertaining to Commercial Mobile Radio Services, Notice of Proposed Rule Making and Notice of Inquiry*, 9 FCC Rcd. 5408, ¶ 3 (1994) (suggesting that the public interest would be served by imposing the same "equal access obligations" on wireless telecommunications licensees as those imposed on wireline licensees, because, *inter alia*, imposition of such obligations would "foster regulatory parity between wireline and wireless services."), *proceeding terminated*, 11 FCC Rcd. 12456 (1996).

particularly when the services being regulated differently come to look more and more the same to consumers."[258] Accordingly, Bush Administration policy now holds that:

> where possible, we should promote competition through a technology-neutral paradigm. . . . [B]roadband services can be deployed over telephony, cable, wireless and satellite platforms. The differing histories and regulations surrounding each type of platform makes absolute regulatory parity difficult to achieve, but it is important to try to regulate comparable services in a manner that does not interfere with marketplace outcomes.[259]

Consistent with this policy, the FCC in the March 15, 2002 *Cable Modem Order & NPRM* expressed a desire "to create a rational framework for the regulation of competing services that are provided via different technologies and network architectures."[260] Noting that "residential high-speed access to the Internet is evolving over multiple electronic platforms, including wireline, cable, terrestrial wireless and satellite," the Commission stated that it "strive[s] to develop an analytical approach that is, to the extent possible, consistent across multiple platforms."[261]

At present, of course, the FCC's regulations that govern the residential market for high-speed Internet access are not consistent across multiple platforms. Rather, cable modem service is currently not subject to any unbundling or open access requirements, while cable's chief rival (DSL) *is* subject to such requirements.[262] Regulatory parity between cable and DSL, of course, could be attained without implementing cable open access: the FCC could reach the same goal by instead choosing to relax the unbundling requirements that currently apply to DSL service, as it has proposed to do.[263] However, implementation of cable open access certainly remains one possible vehicle for the Commission to adopt a technology-neutral paradigm for regulating residential high-speed Internet access that is consistent across multiple platforms.[264]

---

[258] Rosemary Harold, **Cable-Based Internet Access: Exorcising the Ghosts of "Legacy" Regulation,** 28 N. Ky. L. Rev 721, 728 (2001).

[259] Keynote Address by Ass't Sec'y of Commerce Nancy Victory before the Alliance for Public Technology Broadband Symposium, Washington, D.C. (Feb. 8, 2002), *online at* <http://www.ntia.doc.gov/ntiahome/speeches/2002/apt_020802.htm>. *See also, e.g.*, Remarks of FCC Comm'r Michael K. Powell before the Progress & Freedom Foundation (Dec. 8, 2000), *online at* <http://www.fcc.gov/commissioners/powell> (calling "rationaliz[ation]" of regulations governing residential high-speed Internet access, on ground that "a bit is a bit," regardless of the transmission system).

[260] *In re Inquiry Concerning High-Speed Access to the Internet Over Cable and Other Facilities, Declaratory Ruling & NPRM*, FCC 02-77, GEN Docket No. 00-185, 2002 WL 407567, ¶ 6 (rel. Mar. 15, 2002).

[261] *Id.*

[262] *See* Part IV, supra.

The implementation of direct access to INTELSAT, in contrast, had precisely the opposite effect. Rather than introducing a technology-neutral paradigm into the regulation of international telecommunications services, direct access to INTELSAT replaced the rough regulatory parity that had already developed in that market with unique regulatory *dis*parity. Specifically, even before direct access was implemented, COMSAT already was prohibited from bundling the provision of satellite transmission capacity (over which it historically enjoyed market power) with its provision of ancillary services that were subject to competition.[265] Thus, COMSAT, like other dominant retail international common carriers,[266] was long required to provide its services to its retail customers on an unbundled "a la carte" basis.

---

[263] *See Appropriate Framework for Broadband Access to the Internet Over Wireline Facilities*, *Universal Service Obligations of Broadband Providers*, *Notice of Proposed Rulemaking*, FCC 02-42, CC Docket No. 02-33, 2002 WL 252714 (rel. Feb. 15, 2002) (initiating proceeding to consider whether DSL unbundling requirements should be relaxed to attain regulatory parity with cable modem service).

[264] *See In re Inquiry Concerning High-Speed Access to the Internet Over Cable and Other Facilities, Declaratory Ruling & NPRM*, FCC 02-77, GEN Docket No. 00-185, 2002 WL 407567, ¶¶ 77-78 (rel. Mar. 15, 2002) (suggesting that implementing cable open access would advance the FCC's goal of regulatory parity, and querying whether there is any legal or policy reason why cable modem service should not be subject to the same unbundling and access requirements to which DSL service is subject).

[265] *See, e.g.*, *COMSAT Non-Dominant Order*, 13 FCC Rcd. 14083, ¶¶ 169-70 (1998) (COMSAT must unbundle its earth station services from its provision of INTELSAT space segment transmission capacity); *Communications Satellite Corp. Structural Relief Order*, 8 FCC Rcd. 1531, 1536 (1993) (COMSAT must unbundle its maritime value-added services from its provision of ship-to-shore transmission capacity); *Comsat Corp. Petition for Further Partial Waiver of Structural Separation Requirements*, 11 FCC Rcd. 7938, p 29 (1996) (COMSAT must unbundle its provision of mobile terminal equipment, related software, and other maritime value-added services from its provision of ship-to-shore transmission capacity).

[266] *See* 47 C.F.R. §§ 63.10-63.17 (setting forth regulations that govern international common carriers). *Cf.* Phil Weiser, *Paradigm Changes in Telecommunications Regulation*, 71 U. Colo. L. Rev. 819, 827 (2000) ("From an antitrust perspective, mandates to unbundle facilities—whether real or virtual—serve two purposes: first, to prevent the use of market power in one market to disadvantage competition in a second market; and second, to facilitate competitive entry into a market where a company's entrenched monopoly power would be extraordinarily difficult to overcome.").

At the same time, however, until 1999, neither INTELSAT nor any other facilities-based satellite or submarine cable operator had been required to sell raw international telecommunications transmission capacity to an unaffiliated competitor on the same *wholesale* terms and conditions that it sold such capacity to its own retail affiliate.[267] Thus, a satellite operator such as Hughes/PanAmSat has never enjoyed any right to purchase transoceanic submarine cable transmission capacity at wholesale rates from a cable provider such as AT&T. Similarly, a transoceanic submarine cable operator such as AT&T has never enjoyed any right to purchase satellite transmission capacity at wholesale rates from a satellite operator such as Hughes/PanAmSat.

Direct access to INTELSAT carves out a unique exception to this general rule, by imposing a unique regulatory burden on INTELSAT that applies to none of INTELSAT's intermodal or intramodal competitors. Under direct access, INTELSAT is required to supply transmission capacity at wholesale rates to unaffiliated satellite operators, submarine cable operators, or non-facilities-based resellers.[268] None of those competitors, however, are burdened by any corresponding obligation to supply transmission capacity at wholesale rates to INTELSAT, to COMSAT, or to anyone else.

As discussed above, one can reasonably debate whether implementing cable open access would be the most efficacious way for the FCC to advance its goal of a technology-neutral regulatory parity. It is clear, however, that imposing unique direct access requirements on INTELSAT that do not apply to any of its facilities-based head-to-head competitors cannot be said to advance the cause of regulatory parity. Accordingly, the parity-based case for cable open access is comparatively stronger than the parity-based case for "direct access to INTELSAT."

## VII. CONCLUSION

The FCC is now engaged in resolving whether to require cable system operators who provide cable modem service to residential users to furnish cable transmission capacity to unaffiliated Internet Service Providers. To resolve this controversy, the FCC has expressed a desire "to develop an analytical approach that is, to the extent possible, consistent across multiple platforms."[269] This comment may have been intended to specifically highlight the fact that DSL service is currently subject to a panoply of access and unbundling requirements that do not now apply to cable modem service. However, it can also be read more broadly to suggest that in a world of increasing technological convergence and increasing intermodal competition, a more universally consistent analytical approach is needed to resolve the many analogous disputes over competitive access to proprietary bottleneck facilities that arise in a broad range of communications contexts.

---

[267] Until 1991, the FCC actually *limited* the quantity of international telecommunications transmission capacity that a facilities-based provider could supply at wholesale rates to retail resellers. *See* Charles H. Kennedy & M. Veronica Pastor, *An Introduction to International Telecommunications Law* 107 (1996). In 1991, however, the FCC eliminated such limitations after concluding that unlimited resale opportunities would lead to reduced retail prices. *Id.* at 107-08 & n.12 (citing 70 Rad. Reg. 2d (P&F) 160 (1991)).

[268] *See Direct Access To The INTELSAT System*, 14 FCC Rcd 15703 (1999).

The issues raised by the current dispute over "cable open access" are substantially analogous to those raised in the longstanding dispute over "direct access" to the INTELSAT satellite system. That dispute was resolved in 1999, when the FCC authorized unaffiliated competitors to obtain direct access to INTELSAT, on the grounds that such a policy would : (1) encourage the widest possible deployment of communications facilities; (2) encourage competition among providers of communications service; and (3) benefit consumers by facilitating lower prices and more diverse service offerings. Each of these arguments have been raised by proponents of cable open access. In fact, without exception, these criteria each support the implementation of cable open access today at least as strongly as they supported the implementation of direct access to INTELSAT in 1999. Accordingly, implementation of cable open access would be analytically consistent with the implementation of direct access to INTELSAT. Conversely, an FCC decision *not* to implement cable open access would be analytically *inconsistent* (indeed, irreconcilable) its decision to impose INTELSAT direct access decision.

---

[269] *In re Inquiry Concerning High-Speed Access to the Internet Over Cable and Other Facilities, Declaratory Ruling & NPRM*, FCC 02-77, GEN Docket No. 00-185, 2002 WL 407567, ¶ 73 (rel. Mar. 15, 2002), *petitions for review pending sub nom., EarthLink, Inc. v. FCC*, Docket No. 02-1097 (D.C. Cir. filed Mar. 26, 1997).